\documentclass[fleqn,usenatbib]{mnras}

\usepackage{newtxtext,newtxmath}

\usepackage[T1]{fontenc}

\DeclareRobustCommand{\VAN}[3]{#2}
\let\VANthebibliography\thebibliography
\def\thebibliography{\DeclareRobustCommand{\VAN}[3]{##3}\VANthebibliography}

\usepackage{graphicx}	
\usepackage{amsmath}	

\title[Supermassive stars in the first star clusters]{Formation of supermassive stars in the first star clusters}

\author[B. Reinoso et al.]{
Basti\'an Reinoso$^{1}$\thanks{E-mail: bastian.reinoso@uni-heidelberg.de},
Ralf S.\ Klessen$^{1,2}$, 
Dominik Schleicher$^{3}$,
Simon C.~O.\ Glover$^{1}$,
and P. Solar$^{3}$
\\
$^{1}$Universit\"at Heidelberg, Zentrum f\"ur Astronomie, Institut f\"ur Theoretische Astrophysik, Albert-Ueberle-Str. 2, 69120 Heidelberg, Germany \\
$^{2}$Universit\"{a}t Heidelberg, Interdisziplin\"{a}res Zentrum f\"{u}r Wissenschaftliches Rechnen, Im Neuenheimer Feld 205, 69120 Heidelberg, Germany \\
$^{3}$Departamento de Astronom\'ia, Facultad Ciencias F\'isicas y Matem\'aticas, Universidad de Concepci\'on, Av. Esteban Iturra s/n Barrio Universitario,\\ Casilla 160-C, Concepci\'on, Chile 
}

\date{Accepted XXX. Received YYY; in original form ZZZ}

\pubyear{2015}

\begin{document}
\label{firstpage}
\pagerange{\pageref{firstpage}--\pageref{lastpage}}
\maketitle

\begin{abstract}
The formation of supermassive stars is believed to be an essential intermediate step for the formation of the massive black hole seeds that become the supermassive black holes powering the quasars observed in the early Universe. Numerical simulations have shown that supermassive stars can form in atomic-cooling halos when protostars reach accretion rates higher than $\sim10^{-2}$~M$_\odot$~yr$^{-1}$ and fragmentation is suppressed on pc scales. It is however still uncertain if a supermassive star still emerges when fragmentation occurs at smaller scales and a cluster of stars is formed instead. 
In this work we explore the problem of massive object formation due to the interplay of collisions and accretion in star clusters at low metallicity.
We model a small embedded cluster of accreting protostars following sub-parsec scale fragmentation during the collapse of a primordial gas cloud and follow its evolution by performing $N$-body plus hydrodynamical simulations. Our results show that supermassive stars with 10$^3$ and 10$^4$ M$_\odot$ are always formed due to the interplay of collisions and accretion, and in some cases these objects are part of a binary system. The resulting supermassive star is surrounded by tens of smaller stars with typical masses in the range $1$--$100$~M$_\odot$.
\end{abstract}




\section{Introduction}
As of today more than 200 quasars have been detected at redshift
$z>5.7$ \citep{Fan2006,Mortlock2011,Wu15Nature,Banados2018,Reed2019,Onoue2019,Banados2021,Wang21}, with masses larger than 10$^9$~M$_\odot$, and notably, more than 10$^{10}$~M$_\odot$ for SDSS J010013.02+280225.8 \citep{Wu15Nature}.
Explaining the formation and growth of the supermassive black holes (SMBHs) powering those quasars, at an age of the Universe of less than a billion years, is still an important open problem in astrophysics \citep[see the reviews by][]{Volonteri10,review_woods19}.
A natural approach to solve this problem is to find and study the processes capable of yielding massive black holes (BHs) early in the history of the Universe. These early-formed massive BHs are the seeds that grow further by accreting matter, continuous mergers, or both; becoming finally the most distant quasars observed today. 

The most straightforward path that yields massive BH seeds
comes from the death of massive population III stars \citep{Abel2002,Heger2002,Heger2003,Klessen19}, whose initial mass function (IMF) is believed to be top heavy as supported
by recent numerical simulations \citep{Stacy2016,Fraser2017,Rafeel18,Sharda2020}. This scenario however faces important limitations as the formed seeds are still too light ($\lesssim 10^2$~M$_{\odot}$). In addition, because massive Pop~III stars are very effective at expelling gas from the low mass halos in which they form, the black holes formed from them are `born starving' in regions of low gas density and thus cannot grow efficiently by gas accretion \citep{Johnson2007,Smith2018}.
An alternative pathway for massive black hole seed formation is the runaway growth of a single star due to stellar collisions in very dense star clusters \citep{Omukai2008,Katz2015,Sakurai2017,Sakurai2019,Reinoso18,Reinoso20,Vergara21}, or black hole mergers in dense black hole clusters \citep{Davies11,Lupi14}. This channel yields massive objects with typical masses of 10$^3$~M$_\odot$. These BHs could grow to 10$^9$~M$_\odot$ by $z\sim$7 if they accreted continuously at the Eddington limit, but this is an unlikely scenario considering the environment in which those seeds emerge.
A recent work by \cite{Escala21} suggests that a runaway collision process in nuclear star clusters could produce BHs with masses up to 10$^{9}$~M$_\odot$.

The pathway that yields the most massive BH seeds is the so-called direct collapse black hole (DCBH) scenario, and as of today it seems the most plausible explanation for the highest redshift quasars observed.
Initially proposed by \cite{Rees1984}, this formation channel consists of the accumulation of a huge amount of matter in a sufficiently small volume, following the collapse of a pristine gas cloud. This process yields a supermassive star (SMS) that collapses to a BH due to the post-Newtonian instability \citep{Chandrasekhar64}.

Stellar structure calculations show that SMSs are inflated objects, with effective temperatures of ~10$^4$~K, that can reach final masses of ~10$^5$~M$_\odot$ before collapsing due to the post-Newtonian instability \citep{Chandrasekhar64,Hosokawa12,Hosokawa13,Schleicher2013,Woods17,Haemerle18b,Haemerle18a}. Given their low effective temperature, they are unable to produce ionizing photons that may terminate accretion due to radiative feedback. Furthermore, considering the gas-rich environments in which those objects form, they are promising candidates to produce the massive BH seeds that can grow further by mass accretion.

Recent numerical simulations explored the collapse of pristine gas clouds in the early Universe and found that an essential condition for the formation of SMSs in such environments is the suppression of molecular hydrogen cooling, which would otherwise lead to fragmentation of the cloud and the formation of population III stars.
Preventing the cooling due to molecular hydrogen requires a decrease of its abundance by photodetachement of the H$^-$ ion and the destruction of the H$_2$ molecule. This can be achieved in the presence of a strong radiation background that carries photons in the Lyman-Werner bands ($11.2~{\rm eV} \leq h\nu \leq 13.6~{\rm eV}$) and dissociates the H$_2$ molecule, along with infrared photons ($h\nu\geq 0.76$~eV) that lower the abundance of H$^{-}$, a catalyst for H$_2$ formation. This can occur if two pristine halos remain at a small separation such that once star formation begins in one of them, the other is exposed to a high Lyman-Werner radiation intensity, thus suppressing molecular hydrogen cooling. This is termed the `synchronized pairs' scenario \citep{Dijkstra2008,Visbal2014,Chon2018}. Once molecular hydrogen cooling has been suppressed, cooling occurs primarily via collisional excitation of hydrogen atoms, provided that the gas temperature is high enough to make this process efficient. Halos in which this is the case are often referred to as atomic-cooling halo.


The radiation intensity needed to suppress molecular hydrogen cooling is usually expressed in units of $J_{21}$, where $J_{21} = 1$ corresponds to a radiation intensity of $10^{-21}$~erg~cm$^{-2}$~s$^{-1}$~sr$^{-1}$~Hz$^{-1}$ at the Lyman limit \citep[see e.g.][]{Omukai2001,Latif2015}. The required radiation intensity on the atomic-cooling halo could be as high as $J_{21}$=$1\,000$ \citep{Regan2014,Latif2015} or even higher when considering an X-ray background \citep{Inayoshi2015,Glover2016}, and the true value has important implications for the number density of DCBHs \citep{Dijkstra14,Inayoshi2015,Chon2016,Chon2018}.

It has been suggested \citep{Wise2019} that extremely high radiation intensities are not a necessary condition as long as the dark matter halo grows rapidly through mergers. The dynamical heating induced by this period of rapid growth, combined with a moderate Lyman-Werner flux of $J_{21}\sim$~3 can still produce accretion rates of the order of 0.1--1~M$_\odot$~yr$^{-1}$ onto the central object. Once the right conditions are met and high accretion rates achieved ($>0.04$~M$_{\odot}$~yr$^{-1}$), an SMS can still emerge.

The ideal places for the emergence of DCBHs are overdense regions in the early Universe, as they provide intense radiation backgrounds and rapid halo growth. This has been investigated via semi-analytic models by \cite{Lupi2021}, suggesting that the `synchronized pairs' channel as well as the dynamically heated halos can produce several BH seeds in these environments. 

Although a single object forms initially in idealized scenarios of pristine atomic-cooling halos
irradiated by a high intensity LW background, it is important to follow its evolution for longer times in order to place constraints on its final mass. High resolution numerical simulations have shown that fragmentation is unavoidable in the accretion disk for the high accretion rates expected in these environments, and fragmentation is seen on $\sim$au scales \citep{Clark2011, Greif12, Latif2016, Becerra18, Suazo2019, Patrick2020, Wollenberg20, Latif21, Jaura22, Prole22a, Prole22b}. 
It is therefore important to understand the fate of the halos that failed to remain metal-free and/or of the ones in which an important degree of fragmentation is expected.
This scenario is now being explored, and various models, both numerical and analytical, have shown that SMSs with 10$^{4-5}$~M$_\odot$ might still be able to form \citep{Boekholt2018,Alister20,Tagawa20,Das21,Schleicher22}. These results seem to be confirmed by the more sophisticated simulations of \cite{Chon20}. Additionally, \cite{Sassano21} showed that under Eddington-limited accretion, the heavy black hole seeds ($\sim 10^5$~M$_\odot$) are able to produce $10^9$~M$_\odot$ BHs at $z \sim 6$ (see also \citealt{Trinca22} for a similar analysis involving light BH seeds).

In this paper we present a set of $N$-body plus hydrodynamics simulations that include mass accretion, mass-radius parametrizations and stellar mergers to model the central region of a collapsed primordial cloud in which multiple protostars are present.
We explore two environments similar to the ones expected in atomic-cooling halos to assess the impact of fragmentation at sub parsec scales during the assembly of DCBHs. We describe our simulation setup in Sec.~\ref{sec:setup}, then present our results in Sec.~\ref{sec:results} and a discussion in Sec.~\ref{sec:discussion}.



\section{Setup}
\label{sec:setup}
In this section we describe the initial conditions for our models, the numerical codes used, and additional algorithms that we include in our simulations.

\subsection{Initial conditions}
\label{sec:init_cond}
The clusters are modelled to consist of a combination of gas and protostars, the former represented by SPH particles and the latter by particles that interact only through gravity, which we also refer to as $N$-body particles throughout this paper. We model two clusters that differ only in the total mass being $10\,025.6$ in one case and $30\,025.6$~M$_\odot$ in the other. In both cases, we start with a total mass of 25.6~M$_\odot$ in protostars, so the initial gas masses are $10\,000$ and $30\,000$~M$_\odot$, respectively. The initial number of protostars is 256 and each of them has a mass of 0.1~M$_\odot$ which is consistent with the mass of protostars formed in atomic-cooling halos \citep{Becerra2015}.
The gas is sampled with $2^{18}$ SPH particles.
For each set of particles (SPH and protostars), the positions are sampled from a Plummer distribution \citep{Plummer1911} with a Plummer radius $R_{\rm p}\sim 0.077$~pc such that the half-mass radius is $R_{\rm h}\sim1.3R_{\rm p}=0.1$~pc and we impose a cut-off radius of 5 Plummer radii for each model such that all the mass is enclosed within $\sim$0.4~pc. This yields an initial number density for protostars of 956~pc$^{-3}$. We adopt this distribution for simplicity, as the precise distribution of gas and protostars will be unknown. However, it ensures a meaningful behavior of both quantities in the central region, where the density profile will be flat, while the behaviour in the outer parts will approximately resemble the behaviour found in cosmological simulations \citep[e.g.][]{Latif2015}. The velocities of the protostars are obtained by imposing virial equilibrium condition. We relax the Plummer distribution of SPH particles and then inject a spectrum of non-compressive Kolmogorov turbulence with Mach number ${\cal M} = 1$ as found in numerical simulations by \citet{Latif2013}.

\subsection{Numerical simulations}
To run our simulations we use the Astrophsyical MUlti-purpose Software Environment \citep[AMUSE\footnote{https://github.com/amusecode/amuse}, see][]{AMUSE_Portegies09,AMUSE_Portegies13,AMUSE_Pelupessy13,Portegies2018}, a {\small PYTHON} interface designed to couple existing numerical codes, offering great flexibility and allowing us to relatively easily include new algorithms such as mass accretion onto the protostars, sink particle creation, a treatment for stellar collisions, and mass radius relations for the protostars, all of them described in the next subsections.

\subsection{$N$-body-SPH coupling}
We couple the pure $N$-body code {\small PH4} \citep{McMillan96} and the SPH code {\small FI} \citep{Hernquist_Katz1989,Gerritsen1997,Pelupessy2004} by means of the {\small BRIDGE} method \citep{Fujii2007} via the \textit{bridge} class included in {\small AMUSE}. This method consists of calculating the gravitational acceleration at the position of the $N$-body particles using the SPH particles and vice versa, i.e., the particles in one code kick the particles in the other code. To ensure that the coupling does not violate Newton's third law, we use the code {\small FastKick} to perform the kicks, with a constant gravitational smoothing length of $0.5$~au, approximately equal to the smallest smoothing length among all the SPH particles. By doing so we employ the same gravitational smoothing kernel for both sets of particles and make sure that the gravitational forces among them are symmetric.

For evolving the particles in the $N$-body code we use a smoothing length of $1$~R$_\odot$ in order to accurately solve gravitational interactions between the protostars.
We include an external pressure floor in the SPH code by modifying the momentum equation in an analogous way as done in \cite{Benz1990,Clark2011}. This external pressure is equal to the pressure of the cloud at the cut-off radius, and corresponds to $\sim1.72\times 10^{-7}$~g~cm$^{-1}$~s$^{-2}$ for the most massive cloud and $\sim7.45\times 10^{-8}$~g~cm$^{-1}$~s$^{-2}$ for the less massive cloud. The external pressure is required to stabilize the clouds against vacuum boundary conditions. 

Finally, we modified the code {\small FI} to include a modified equation of state of the form
\begin{equation}
\label{eq:EOS}
    T = T_0 \left[ 1 + \left( \frac{\rho}{\rho_{\rm c}} \right)^{\gamma - 1} \right],
\end{equation}
so that the gas behaves isothermally, with a temperature $T_0=8\,000$~K at low densities, but becomes adiabatic at densities above $\rho_{\rm c}=10^{15}$~cm$^{-3}$, as found in 1D and 3D models including detailed chemical networks \citep{Omukai2008,Becerra2015}.
We use an adiabatic index $\gamma=5/3$.\\


The treatment of stellar collisions (described in Sec.\ref{sec:stellar_collisions}) as well as the mass radius relations are implemented at the {\small PYTHON} level. The mass accretion (see Sec.\ref{sec:gas_accretion}) and sink particle creation (see Sec.\ref{sec:sink_particles}) algorithms are written in {\small FORTRAN} and included as {\small PYTHON} functions via {\small F2PY} for adequate performance. This offers the advantage of easily replacing any of the codes used without having to re-write these routines.

The time integration consists sequentially of the Kick-Drift-Kick (KDK) integration with \textit{bridge} during which stellar collisions and sink particle creation are solved, followed by the computation of accretion onto the protostars, and the treatment of stellar ejections. We impose a maximum timestep of 5~yr for \textit{bridge}
in order to perform the accretion steps rather frequently, given the rapid accretion rates expected in this environment.

The densest gas is typically found around accreting protostars and thus it is often accreted after every accretion step. Nevertheless there are regions where the gas
becomes very dense and thus the associated free-fall time can be of the order of 10$^{-4}$~yr. Because of this we implemented an adaptive time-stepping algorithm for \textit{bridge}, such that the timestep is reduced by factors of 2 until becoming smaller than the shortest free-fall time.
The timestep can increase by a factor of 2 after each timestep only if the shortest free-fall time is more than twice the current timestep.


\subsection{Protostars and stars}
The $N$-body particles in our simulations are meant to represent, in an approximate way, protostars and stars, but they interact here only through gravity and we do not include any type of feedback.
By taking advantage of the particle sets provided by {\small AMUSE} we assign additional properties to our $N$-body particles such as: \textit{stage} and \textit{luminosity}. The \textit{stage} property indicates if an $N$-body particle has already entered the main sequence phase or if it is still in the protostellar phase. This distinction is important as inflated protostars can still contract to the main-sequence phase and thus will follow different mass-radius relations. A correct determination of the size of the particles is an essential feature needed in this very dense collisional environment. 
For the determination of the radius of each $N$-body particle
we incorporate new {\small PYTHON} functions without the need to modify any code. The mass radius parametrization that we use is briefly described in Sec.~\ref{sec:m-r_param} and further details are provided in Appendix~\ref{sec:app_mr}.

\subsection{Mass-radius parametrization}
\label{sec:m-r_param}
We use a parametrization of the mass-radius relation based on the works by \cite{Hosokawa09} and \cite{Hosokawa12,Hosokawa13}. This implementation is simplified by the use of the \textit{stage} property for our $N-$body particles. We define three stages in which an $N$-body particle can be. The possible stages are the \textit{protostar}, \textit{star}, and \textit{supermassive star} stages. 

\begin{figure}
	\includegraphics[width=\columnwidth]{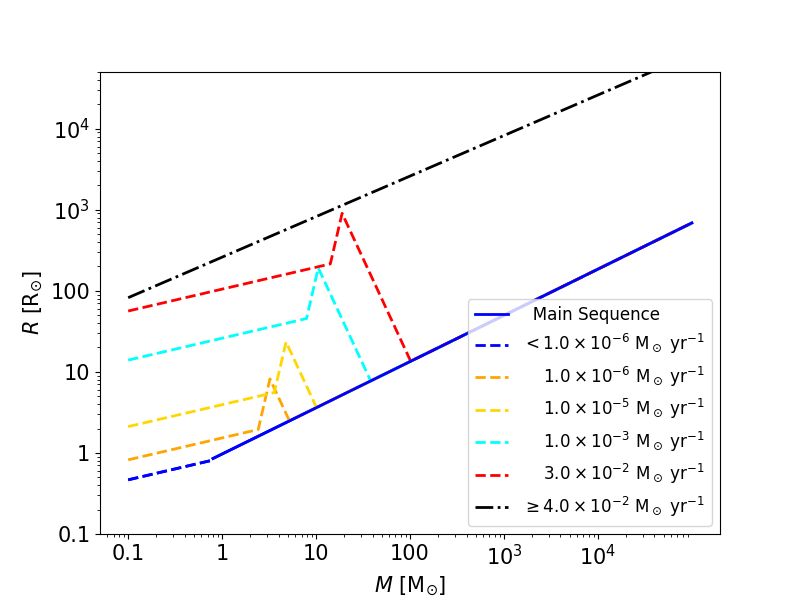}
    \caption{The adopted mass radius relations for accreting protostars in our models.}
    \label{fig:MR_relations}
\end{figure}

For a particle in the \textit{protostar} stage, the mass-radius relation depends on its mass and accretion rate. We present in Fig.~\ref{fig:MR_relations}, with dashed lines, the mass radius relations for protostars accreting at different rates. The implementation is described in Appendix~\ref{sec:app_mr}.

For a particle in the \textit{star} stage, the mass-radius relation is given by:
\begin{equation} 
 R_*=0.97 \left(\frac{M_*}{{\rm M_\odot}}\right)^{0.57} \rm R_\odot.
\end{equation}
It is used once the protostars enter in the main-sequence stage, and corresponds to the blue solid line onto which most of the other lines converge in Fig.~\ref{fig:MR_relations}.

An $N$-body particle enters into the \textit{supermassive star} stage if it is still on the \textit{protostar} stage and if its accretion rate becomes larger than a critical accretion rate $\dot{M}_{\rm crit}$. The mass-radius relation for these particles is given by:
\begin{equation}
    \label{eq:M_R_SMS}
    R_* = 2\,600 \left( \frac{M_*}{100~{\rm M_\odot}}\right)^{1/2} {\rm R_\odot}.    
\end{equation}
It is shown with a dot-dashed line in Fig.~\ref{fig:MR_relations}.

We note that an $N$-body particle in the \textit{protostar} stage will eventually contract to the main sequence after becoming massive enough. 
Similarly, an $N$-body particle in the \textit{supermassive star} stage can also contract to the main sequence if its accretion rate drops below a critical value $\dot{M}_{\rm crit}$ for a time longer than the Kelvin-Helmholtz (KH) timescale $t_{\rm KH}$. We adopt here $\dot{M}_{\rm crit}= 0.04$~M$_\odot$~yr$^{-1}$ \citep{Hosokawa13}.

Due to the structure of supermassive stars, the relevant timescale for contraction is the Kelvin-Helmholtz timescale evaluated at the stellar surface $t_{\rm KH, surf}$
\citep{Schleicher2013,Sakurai2015}.  The value of the KH timescale at the surface for these objects can be approximated as $10$--$100~t_{\rm KH}$ \citep{Sakurai2015}. We explore here these two extreme values, namely $t_{\rm KH, surf}=10~t_{\rm KH}$ and $100~t_{\rm KH}$. The first case is the most pessimistic case for stellar collisions to occur whereas the second is the most optimistic case. In consequence, in our simulations, an $N$-body particle in the \textit{supermassive star} stage will contract if its accretion rate falls below $\dot{M}_{\rm crit}$ for a time longer than $t_{\rm KH, surf}$ given by:
\begin{equation}
t_{\rm KH, surf}= {\rm X}\ {t}_{\rm KH} = {\rm X} \frac{GM^2}{RL},
\end{equation}
with ${\rm X}=10$, or ${\rm X}=100$.

In order to calculate $t_{\rm KH, surf}$ we make use of the \textit{luminosity} property of our $N$-body particles. Luminosites are calculated based on the works by \cite{Hosokawa09} and \citet{Hosokawa12,Hosokawa13} as described in Appendix~\ref{sec:app_mr}.

Finally we also assume that a stellar merger will perturb the new object in such a way that the resulting object has to start over the relaxation process. This means that stellar mergers help to keep the protostars inflated in our simulations.

\subsection{Gas accretion}
\label{sec:gas_accretion}
After every KDK step, we calculate the gas accretion onto the protostars.
We model the gas accretion by employing the algorithm developed by \cite{Hubber13}. For every $N$-body particle we define a spherical volume called the interaction zone, with radius $R_{\rm I.Z.}$. Inside this region the weighted average gas flux onto the central point mass is computed, with the weight calculated via a cubic spline kernel function.

The radius of the interaction zone is adjusted iteratively before every accretion calculation, with a maximum of 50 iterations per particle, to maintain a constant gas mass of $M_{\rm int, max}=50$~$M_{\rm gas}/{N_{\rm SPH}}$ (the mass corresponding to 50 SPH particles). We impose lower and upper limits to the radius of the interaction zone of $r_{\rm min}={\rm max}(10~{\rm au},2R_{\rm star})$ and $r_{\rm max}=500$~au, respectively, so that the interaction zone cannot be smaller than the protostars.  We also impose the angular momentum conservation condition for the accreted gas and the angular momentum feedback from the point particle as described in \cite
{Hubber13}, as well as the prescriptions for spherical or disk accretion.

Once the accretion step has been completed, we update the radius of each star particle according to the mass-radius parametrization described in Sec.~\ref{sec:m-r_param} and in Appendix~\ref{sec:app_mr}.

\subsection{Sink particles}
\label{sec:sink_particles}
Given the high densities reached in our simulations, we decided to include sink particle creation to avoid prohibitively small timesteps. Based on the work by \cite{Hubber13}, specifically on their `NewSink' algorithm, we create sink particles whenever an SPH particle reaches a density higher than 10$^{16}$~cm$^{-3}$ \citep{Becerra2015}, it sits in a minimum of the gravitational potential among its neighbours, it does not overlap existing sinks, and fulfills the density criterion:
\begin{equation}
    \rho_{i} > \rho_{\rm Hill} \equiv \frac{3 X_{\rm Hill} (-\Delta\mathbf{r}_{is'} \cdot \Delta\mathbf{a}_{is'} )}{4\pi G |\Delta\mathbf{r}_{is'}|^2},
\end{equation}
for all existing sinks $s'$ for a given SPH particle $i$. Here $\Delta\mathbf{r}_{is'}$ and $\Delta\mathbf{a}_{is'}$ are the relative position and acceleration of sink candidate $i$ with respect to existing sink $s'$.
We set $X_{\rm Hill}=4$. This Hill sphere criterion ensures that an SPH particle turns into a sink particle in the vicinity of another sink only if the density peak dominates the local gravitational potential.
Once the previous conditions are fulfilled, we remove the SPH particle from {\small FI} and insert a new $N$-body particle in {\small PH4}. The mass, position, and velocity of the new particle are the same as the ones of the removed SPH particle.
The radius of the protostar is initialized to 0.1~R$_\odot$ but is recalculated after every accretion step according to the mass-radius parametrization described in Section~\ref{sec:m-r_param}.\\

The Jeans mass scales as 
\begin{equation}
 M_{\rm J} \propto \left( \frac{T^3}{n}\right)^{1/2},\end{equation}
with $n$ being the number density of the gas. In our simulations the minimum Jeans mass is $\sim$3.96~M$_\odot$. The mass resolution is equal to twice the mass contained inside the smoothing length of an SPH particle. In our case this is
\begin{equation}
    M_{\rm res} = 2N_{\rm neigh} \frac{M_{\rm gas}}{N_{\rm SPH} },
\end{equation}
where $N_{\rm neigh}=64$ is the number of neighbours for one SPH particle as adopted in the code {\small FI}, $M_{\rm gas}$ is the initial cloud mass, and $N_{\rm SPH}$ is the number of SPH particles.

In order to avoid artificial fragmentation, our
simulations need to resolve the Jeans mass \citep{BateBurkert1997}. We achieve the mass resolution by using $N_{\rm SPH}=1\,048\,576$ for all our simulations.

\subsection{Ejections}
The $N$-body particles can also be ejected from the cluster. A particle is considered to have been ejected once it fulfills three criteria: its distance to the centre of mass of the system is $\geq1.4$~pc; its gravitational potential energy plus kinetic energy per unit mass is positive; and it is moving away from the cluster, i.e., $\textbf{r} \cdot \textbf{v}>\textbf{0}$. Ejected particles are removed from the simulation.

\subsection{Collisions}
\label{sec:stellar_collisions}
A collision between two particles occurs once the radii of two $N$-body particles overlap during the $N$-body integration. This is implemented in {\small AMUSE} with the help of the \textit{stopping conditions}. We activate the \textit{stopping condition} that detects the overlap of two particle's radii in {\small PH4}, i.e., a collision occurs when
\begin{equation*}
    d \leq R_1 + R_2,
\end{equation*}
where $d$ is the separation between the particles and $R_1$ and $R_2$ are their radii.
Once the condition is fulfilled, the integration is interrupted, and then, by implementation at the {\small PYTHON} level, we replace the overlapping particles by a new particle that is placed at the centre of mass of the previous configuration and the new velocity is calculated assuming linear momentum conservation. We assume no mass loss to occur, and thus the new mass is the sum of the masses of the colliding particles, i.e.
\begin{equation*}
M_{\rm new} = M_1 + M_2.
\end{equation*}

In order to determine the new radius of the merger product,
we first determine the stage in which the new particle will be, according to the outcomes shown in Tab.~\ref{tab:new_stage}; and the new track, in case the resulting stage is \textit{protostar}, is assigned as shown in Table \ref{tab:new_track}. Subsequently the new radius is obtained from the corresponding mass-radius relation. Further details are provided in Appendix~\ref{sec:properties_merger}.
After determining the evolutionary stage of the particle, the luminosity is calculated as explained in Appendix~\ref{sec:app_mr}.


\section{Results}
\begin{table*}
	\centering
	\caption{Summary of simulation outcomes. We present for each simulation the initial gas mass, the final time, the quiescent time adopted for contraction to the main sequence for supermassive stars, the simulation outcome, the total accreted mass, the final stellar mass bound to the most massive object, the mass of the most massive object, the efficiency of massive object formation, the total mass in ejected stars, the number of stars bound to the MMO, the number of ejections and the number of collisions.}
	\label{tab:m1e4_outcomes}
	\begin{tabular}{lcccrcclrrrrr} 
		\hline
		Simulation  & $M_{\rm gas}$ & $t_{\rm end}$  & $t_{\rm KH, surf}$& outcome & $M_{\rm accreted}$ & $M_{\rm stellar, bound}$& $M_{\rm MMO}$ & $\epsilon$ & $M_{\rm ejected}$ & $N_{\rm stars}$ &$N_{\rm ejections}$ & $N_{\rm col}$\\
		 & [M$_\odot$] & [yr] & [$t_{\rm KH}$] & & [M$_\odot$] & [M$_\odot$] & [M$_\odot$] & & [M$_\odot$] & & & \\
		\hline
		M1\_t100\_1 & $10^4$ & $200\, 015$ & 100 & single & $5\, 414$ & $5\, 305$ & $5\, 197$ & 0.52 &109 & 56 & 70 & 256\\
		M1\_t100\_2 & $10^4$ & $200\, 005$ & 100 & single & $3\, 815$ & $3\, 482$ & $3\, 311$ & 0.33 & 333 & 62 & 112 & 288 \\
        M1\_t100\_3 & $10^4$ & $200\, 043$ & 100 & single & $4\, 709$ & $4\, 539$ & $3\, 893$ & 0.39 & 170 & 56 & 77 & 341 \\
        M1\_t100\_4 & $10^4$ & $200\, 023$ & 100 & single & $3\, 730$ & $3\, 315$ & $3\, 048$ & 0.30 & 415 & 48 & 153 & 291 \\
        M1\_t100\_5 & $10^4$ & $200\, 017$  & 100 & binary & $5\, 854$ & $4\, 821$  & $4\,096$ & 0.41 & $1\, 033$ & 6 & 141 & 369\\
        M1\_t100\_6 & $10^4$ & $200\,050$ & 100 & binary  & $4\,150$ & $3\,326$ & $2\,831$ & 0.28 & 824 & 15 & 196 & 300\\
        \\
        M1\_t10\_1 & $10^4$ & $200\, 024$ & 10 & single & $5\, 397$ & $4\, 952$ & $4\, 326$ & 0.43 & 445 & 66 & 83 & 343 \\
        M1\_t10\_2 & $10^4$ & $120\, 045$ & 10 & single & $4\, 548$ & $4\, 377$ & $4\, 156$ & 0.42 & 171 & 88 & 79 & 375 \\
        M1\_t10\_3 & $10^4$ & $200\, 022$ & 10 & binary & $6\, 057$ & $5\, 297$ & $4\, 064$ & 0.41 & 760 & 65 & 147 & 364\\
        M1\_t10\_4 & $10^4$ & $200\,029$ & 10 & binary & $5\,262$  & $4\, 468$ & $2\,901$ & 0.29 & $794$ & 86 & 141 & 456 \\
        M1\_t10\_5 & $10^4$ & $200\,036$ & 10 & single & $6\,804$ & $6\,256$ & $4\,858$ & 0.49 & 548 & 89 & 53 & 412 \\
        M1\_t10\_6 & $10^4$ & $112\,701$ & 10 & single & $4\,617$ & $4\,255$ & $4\,135$ & 0.41 &  362 & 56 & 119 & 301 \\
        \\
        M3\_t100\_1 & $3\times10^4$& $200\, 021$& 100 & single & $26\, 108$ & $25\, 808$ & $24\, 418$ & 0.81 & 300 & 13 & 42 & $1\, 892$ \\
        M3\_t100\_2 & $3\times10^4$& $200\, 043$& 100 & single & $26\, 939$ & $26\, 898$ & $26\, 890$ & 0.90 & 41 & 10 & 19 & $1\, 842$  \\
        M3\_t100\_3 & $3\times10^4$& $200\, 009$& 100 & single & $26\, 388$ & $26\, 211$ & $24\, 577$ & 0.82 & 177 & 11 & 34 & $2\, 547$ \\
        M3\_t100\_4 & $3\times10^4$& $200\, 038$& 100 & single & $23\, 312$ & $22\, 850$ & $20\, 365$ & 0.68 & 462 & 36 & 53 & $1\, 844$  \\
        M3\_t100\_5 & $3\times10^4$& $200\, 034$ & 100 & single & $23\, 070$ & $22\, 973$ & $22\, 618$ & 0.75 & 97 & 12 & 50 & $2\, 215$ \\
        M3\_t100\_6 & $3\times10^4$& $200\, 035$ & 100 & single & $26\, 966$ & $26\, 851$ & $24\, 375$ & 0.81 & 115 & 3 & 29 & $2\, 522$ \\
        \\
        M3\_t10\_1 & $3\times10^4$& $200\, 008$& 10 & single & $20\, 981$ & $20\, 831$ & $20\, 435$ & 0.68 & 150 & 13 & 70 & $2\, 283$ \\
        M3\_t10\_2 & $3\times10^4$& $200\, 026$ & 10 & single & $23\, 451$ & $23\, 063$ & $20\, 776$ & 0.69 & 388 & 32 & 61 & $1\, 807$ \\
        M3\_t10\_3 & $3\times10^4$& $200\, 048$ & 10 & single & $25\, 871$ & $25\, 413$ & $22\, 267$ & 0.74 & 458 & 6 & 50 & $2\, 354$ \\
        M3\_t10\_4 & $3\times10^4$& $200\, 014$ & 10 & single & $22\, 585$ & $21\, 889$ & $21\, 733$ & 0.72 & 696 & 10 & 89 & $2\, 445$ \\
        M3\_t10\_5 & $3\times10^4$&  $200\, 039$ & 10 & single & $20\, 778$ & $20\, 481$ & $20\, 368$ & 0.68 & 297 & 6 & 96 & $2\, 297$  \\
        M3\_t10\_6 & $3\times10^4$& $200\,011$ & 10 & single & $27\, 051$ &  $26\, 846$ & $26\, 746$ & 0.89 & $205$ & 7 & 42 & $3\,514$ \\
        
		\hline
	\end{tabular}
\end{table*}

\label{sec:results}
In this section we describe the results obtained from our simulations. We begin by describing the general behaviour of the simulated systems in which we set $t_{\rm KH, surf}=100$~$t_{\rm KH}$ (see Sec.~\ref{sec:m-r_param}). We describe the emergence of massive objects and characterize the final state of the clusters.
We do so first for the simulations with $M_{\rm gas}=10^4$~M$_\odot$ and then for simulations with $M_{\rm gas}=3\times10^4$~M$_\odot$. We then show the impact of setting $t_{\rm KH, surf}=10$~$t_{\rm KH}$ on the final masses of the most massive objects. All the uncertainties reported correspond to the one sigma interval assuming a normal distribution.

\subsection{Clusters with $M_{\rm gas} = 10^4$~M$_\odot$}

\subsubsection{Cluster evolution}
\begin{figure}
	\includegraphics[width=\columnwidth]{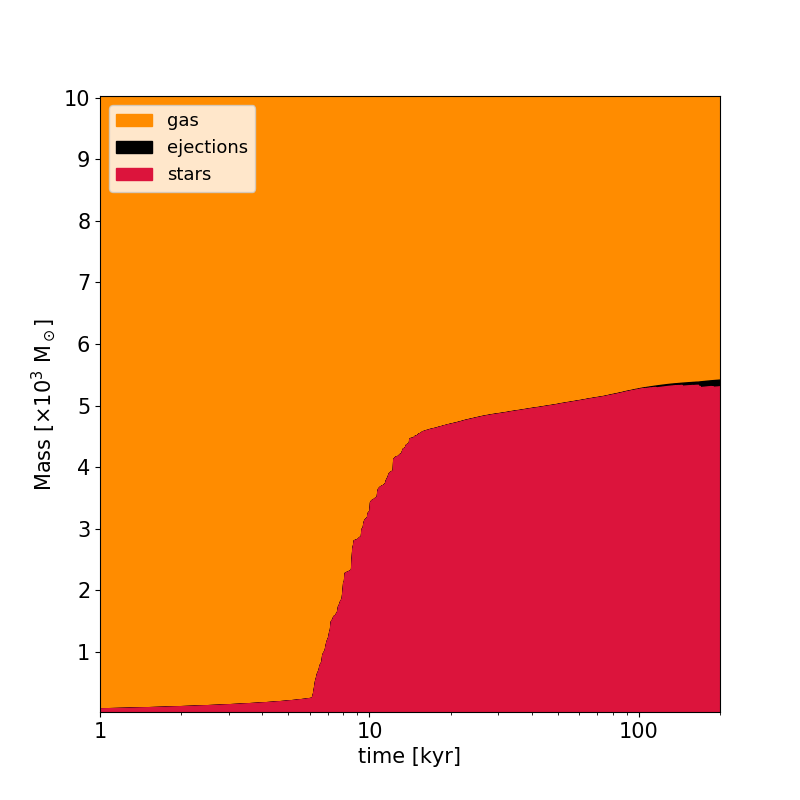}
    \caption{Mass of the gas, bound stars and ejected stars during the evolution of the system for one of our simulations with $M_{\rm gas}=10^4$~M$_\odot$ (M1\_t100\_1).}
    \label{fig:mass_evol_cluster}
\end{figure}

By the time that we stop our simulations, most of the gas accretion has already occurred as depicted in Fig.~\ref{fig:mass_evol_cluster}. The initial phase of the evolution is marked by a contraction of the inner parts of the gas cloud on a free-fall timescale.
We see the contraction of the inner parts of the cloud, up to the radius at which 25 per cent of the total mass is enclosed, i.e. the 25 per cent Lagrangian radius. This takes place at around $6\, 000$~yr and marks the beginning of the short time-span during which most of the gas accretion takes place.

During this contraction, turbulence does not significantly affect the cloud evolution as no substructures appear and we observe a spherical collapse, which in turn causes high accretion rates onto a central object around which a gaseous disk is formed.
This object quickly becomes the most massive object (MMO) of the cluster. 

By the end of the simulations the total accreted mass reaches typical values of $4\, 612\pm798$~M$_\odot$, i.e., $\sim46\pm8$ per cent of the initial mass, and the mean total mass in stars that are still bound to the cluster is $4\, 131\pm791$~M$_\odot$. The mean mass in ejected stars is $481\pm338$~M$_\odot$.

\subsubsection{Emergence of very massive objects}
In most of our simulations we see the formation of a single object that contains most of the accreted mass. The typical mass of the MMO is $3\, 729\pm792$~M$_\odot$. We define a parameter $\epsilon$ to asses the efficiency of the formation of a massive object. It is defined as $\epsilon={M_{\rm MMO}/M_{\rm gas}}$ and shown for each simulation in Table~\ref{tab:m1e4_outcomes}. We find a mean efficiency $\epsilon=0.37\pm0.08$.

We also see that in one third of the simulations, the MMO is in a binary system with another very massive object. We define a pair of bound stars to be in a binary system if they follow a Keplerian orbit and the mass ratio $q=M_1/M_2$ is less than 7. This choice for this mass ratio is arbitrary but allow us to select high mass stars that are in a binary system with the MMO which are the binary systems in which we are interested.\\

We show the properties of the binary systems in Table~\ref{tab:bins_m1e4}, and we note that in simulations with a binary outcome, there are fewer stars in the final stellar system due to more collisions occurring and more ejections due to three body interactions (see columns 11, 12 and 13 in Table~\ref{tab:m1e4_outcomes}).

\begin{figure}
	\includegraphics[width=\columnwidth]{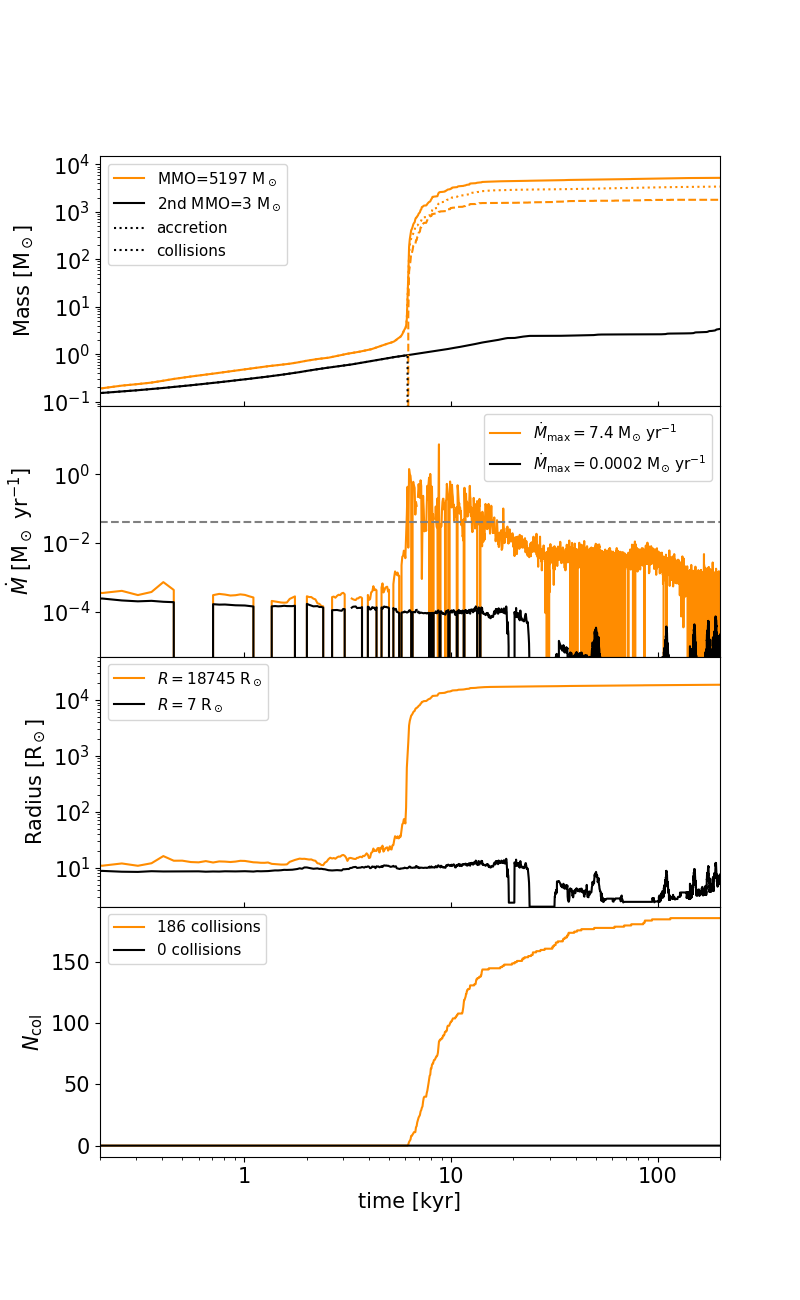}
    \caption{Evolution of the mass $M$, accretion rate $\dot{M}$ and radius $R$ for the MMO and second MMO in one of our simulations (M1\_t100\_2), along with the number of collisions $N_{\rm col}$ as functions of time.}
    \label{fig:MMO_evol_m1e4}
\end{figure}

\begin{figure}
	\includegraphics[width=\columnwidth]{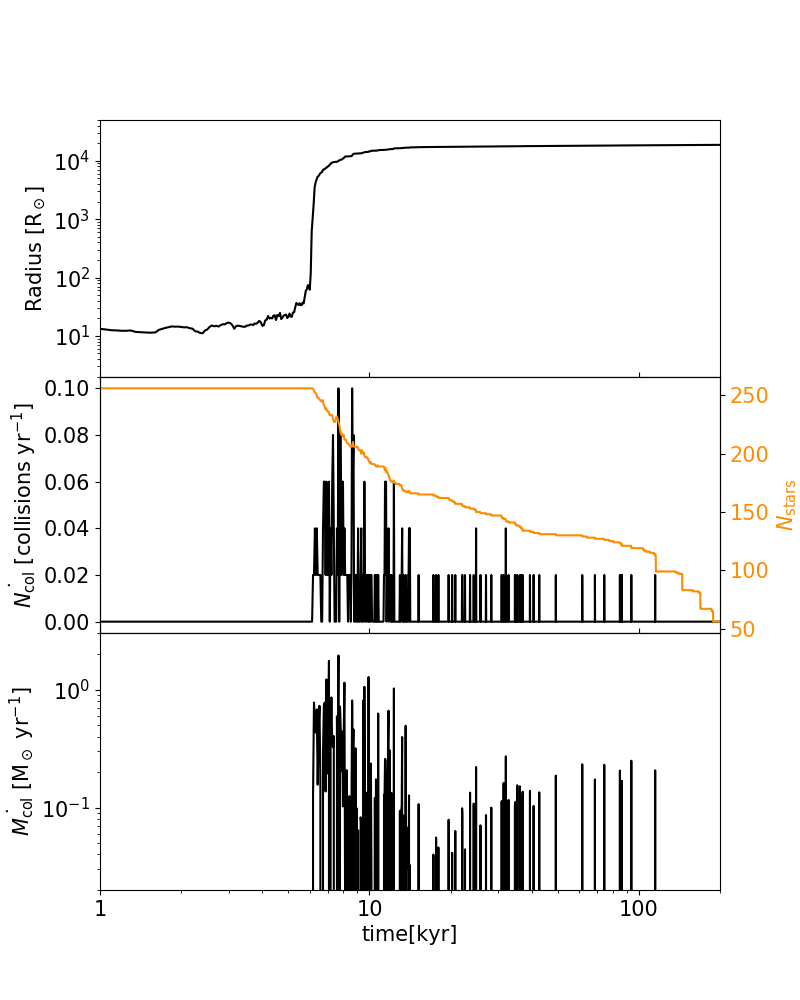}
    \caption{Radius of the MMO, collision rate along with the number of protostars, and mass accretion rate due to collisions as functions of time for one of our simulations (M1\_t100\_1).}
    \label{fig:col_rate_m1e4}
\end{figure}

The overall contraction of the gas cloud causes a strong inflow and therefore a high accretion rate onto one of the central protostars. We show the evolution of this object that becomes the MMO in Fig.~\ref{fig:MMO_evol_m1e4}. The maximum accretion rates in the simulations are a few M$_\odot$~yr$^{-1}$, surpassing during some time the critical accretion rate of $\dot{M}_{\rm crit}=0.04$~M$_{\odot}$~yr$^{-1}$, and thus creating an MMO that emerges in the cloud centre and evolves as a supermassive star. This moment can be distinguished in the second panel of Fig.~\ref{fig:MMO_evol_m1e4} at the point when the solid orange line first crosses the gray dashed line. Because of this the star inflates up to around $2\times10^4$~R$_\odot$, i.e., $\sim93$~au as it now follows the mass-radius relation shown with a black dot-dashed line in Fig.~\ref{fig:MR_relations}. The increased cross section of the central star results in a period of runaway collisions with the MMO. The collision rate reaches a maximum of 0.1 collisions per year just after the MMO has inflated in radius, but starts to decline as the number of protostars decreases, as shown in the middle panel of Fig.~\ref{fig:col_rate_m1e4}. The mass accretion rate due to collisions can reach very high values of up to 2~M$_\odot$~yr$^{-1}$.

\begin{figure}
	\includegraphics[width=\columnwidth]{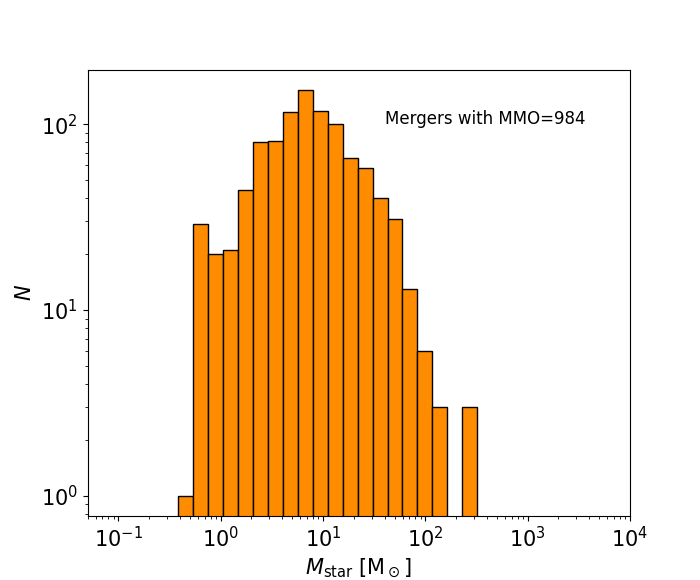}
    \caption{Mass distribution of the particles that merge with the MMO for simulation M1\_t100\_1--6.}
    \label{fig:masses_mergers_mmo_m1e4_100}
\end{figure}

Additionally we see that huge gas densities ($\rho\sim10^{-8}$~g~cm$^{-3}$) around the MMO trigger the formation of $\sim217$ new sink particles on average. Almost all ($95.4\pm 1.7$ per cent) of these new sinks merge with other objects, but only $34\pm7$ per cent of them merge with the MMO. We show the mass distribution of the particles that merge with the MMO in Fig.~\ref{fig:masses_mergers_mmo_m1e4_100}.

\begin{figure}
	\includegraphics[width=\columnwidth]{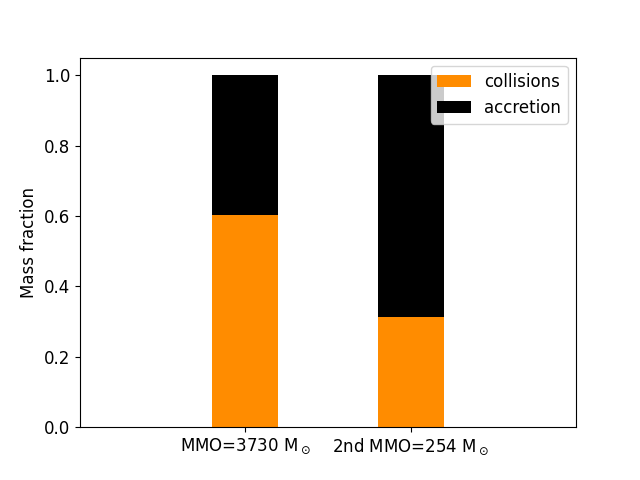}
    \caption{Average mass fraction gained through collisions and accretion, and average masses for the MMO and second MMO for simulations M1\_t100\_1--6.}
    \label{fig:frac_coll_acc_m1e4}
\end{figure}

Despite the accretion rate falling below the critical accretion rate $\dot{M}_{\rm crit}$, the frequent stellar collisions prevent the contraction of the MMO. We note that the mass contributed by collisions to this object is around 60 per cent of its final mass as shown in Fig.~\ref{fig:frac_coll_acc_m1e4}.

\subsubsection{Final cluster properties}
The typical outcome of our simulations is a small cluster of stars with almost no gas left. 
This final stellar cluster in most simulations is made up of $\sim 50$ stars with typical masses in the range $1$--$10~$M$_\odot$ surrounding the MMO. No more significant gas accretion is taking place at 200~kyr, and we would expect radiative feedback from the stars to efficiently evaporate the remaining gas. The final mass functions are similar in shape as well as the number of remaining and ejected stars, although a few clusters contain a binary system and fewer stars remain bound due to the increased number of collisions and three body interactions effectively ejecting lower mass objects.

\begin{figure}
	\includegraphics[width=\columnwidth]{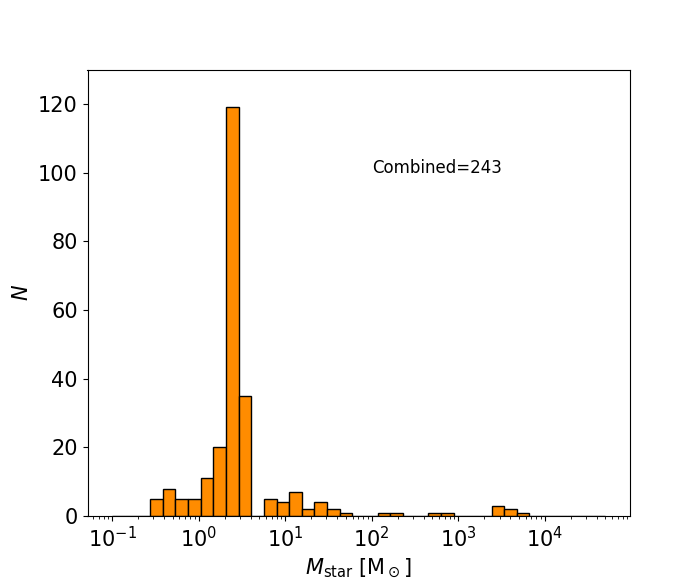}
    \caption{Combined mass distribution for stars bound to the cluster at the end of simulations M1\_t100\_1--6.}
    \label{fig:mass_dist_m1e4}
\end{figure}

\begin{figure}
	\includegraphics[width=\columnwidth]{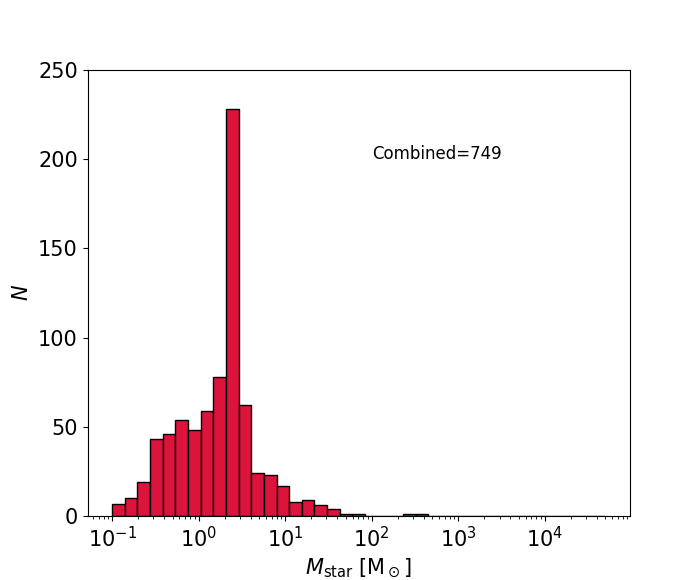}
    \caption{Combined mass distribution of ejected stars at the end of simulations M1\_t100\_1--6.}
    \label{fig:mass_dist_esc_m1e4}
\end{figure}

We present the combined mass distribution at the end of simulations M1\_t100\_1--6 in Fig.~\ref{fig:mass_dist_m1e4}, and the combined mass distribution of ejected particles in Fig.~\ref{fig:mass_dist_esc_m1e4}. Individual mass distributions of bound and ejected particles for each simulation are presented in Figs.~\ref{fig:M_dist_m1e4_individual}~and~\ref{fig:M_dist_esc_m1e4_individual}.

\begin{table}
	\centering
	\caption{Properties of binary systems. We present the mass of the most and less massive object $M_1$ and $M_2$ respectively, the semi-major axis $a$ and the eccentricity $e$. }
	\label{tab:bins_m1e4}
	\begin{tabular}{l r r c c } 
		\hline
		Simulation & $M_1$ & $M_2$ & $a$ & $e$ \\
                   & [M$_\odot$] & [M$_\odot$] & [au] & \\
		\hline
M1\_t100\_5 & $4\,096$ & 688 & 355 & 0.126 \\
M1\_t100\_6 & $2\,831$ & 464 & 240 & 0.375 \\
M1\_t10\_3  & $4\,064$ & 831 & 472 & 0.394 \\
M1\_t10\_4  & $2\,901$ & $1\,224$ & 120 & 0.077\\

		\hline
	\end{tabular}
\end{table}

\subsection{Clusters with $M_{\rm gas}=3\times 10^4$~M$_\odot$}
In this section we describe the general evolution of the clusters with $3\times10^4$~M$_\odot$ in gas, and mention the differences with the less massive clusters.
\subsubsection{Cluster evolution}
\begin{figure}
	\includegraphics[width=\columnwidth]{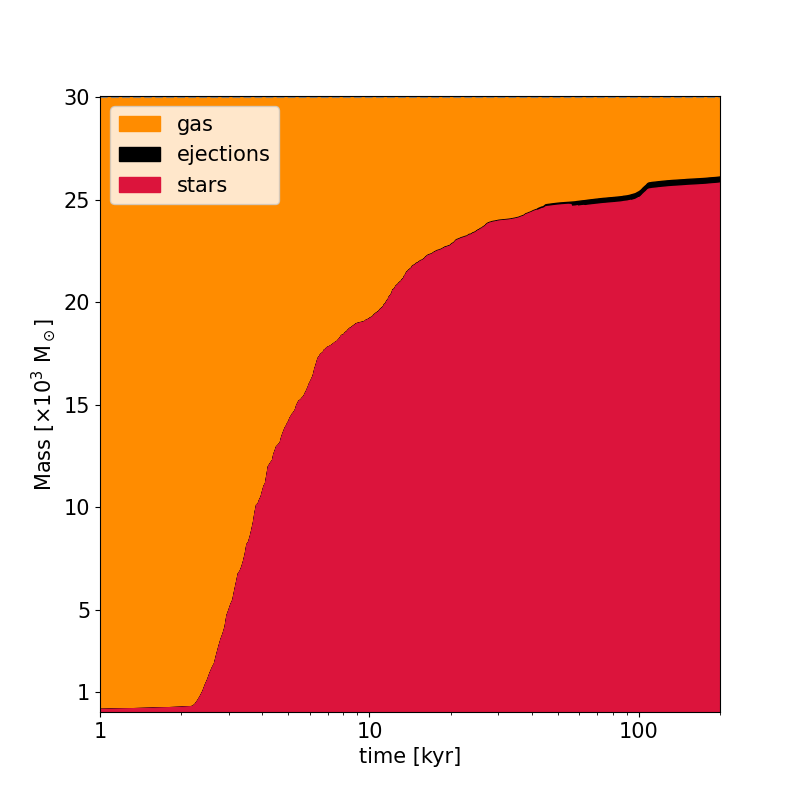}
    \caption{Same as Fig.~\ref{fig:mass_evol_cluster} but for a cluster with $M_{\rm gas}=3\times 10^4$~M$_\odot$.}
    \label{fig:mass_evol_cluster_m3e4}
\end{figure}

The initial behaviour of the gas cloud is the same for all the simulations, and also very similar to the behaviour in the less massive clusters. We see that most of the gas is accreted early on in the cloud evolution as depicted in Fig.~\ref{fig:mass_evol_cluster_m3e4}. During the initial evolution the inner parts of the cloud experience an overall contraction. Specifically, we see a contraction of the 25 per cent Lagrangian radius, which leads to a rapid inflow of gas to the central parts of the cluster in a free-fall time, i.e., $\sim3\, 000$~yr. Unlike in the less massive clusters, we also see a contraction at the 50 per cent Lagrangian radius.

Turbulence seems to have a negligible role here as no substructure appears during the initial contraction and a spherical collapse proceeds.
During the rapid mass inflow either a central object starts to accrete most of the mass, or a new sink particle is created at the centre due to the high gas densities. This central particle reaches accretion rates of several 10~M$_\odot$~yr$^{-1}$, and the average efficiency $\epsilon=0.80\pm0.07$ means that this single object gathers on average $80\pm7$ per cent of the total mass of the cloud. 

\subsubsection{Emergence of very massive objects}
\begin{figure}
	\includegraphics[width=\columnwidth]{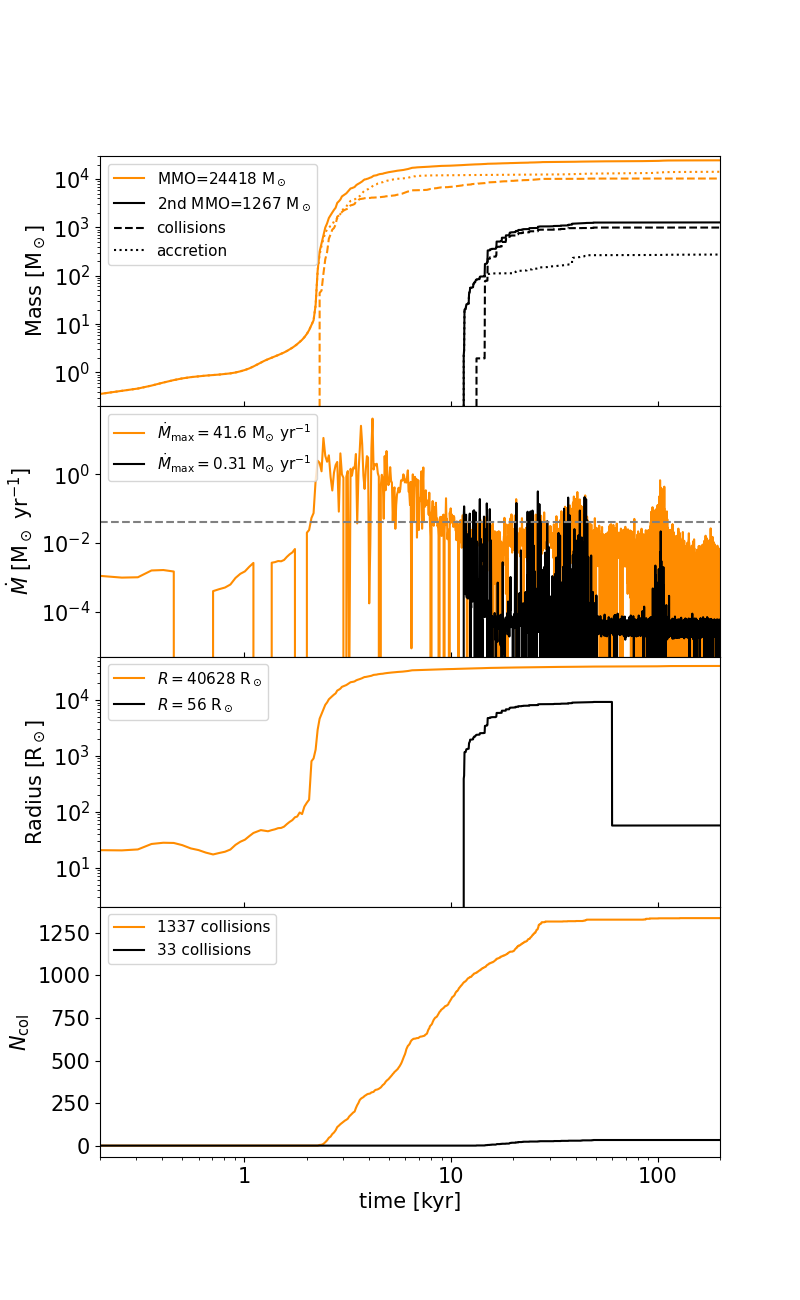}
    \caption{Same as Fig.~\ref{fig:MMO_evol_m1e4} but for a cluster with $M_{\rm gas}=3\times 10^4$~M$_\odot$.}
    \label{fig:MMO_evol_m3e4}
\end{figure}

\begin{figure}
	\includegraphics[width=\columnwidth]{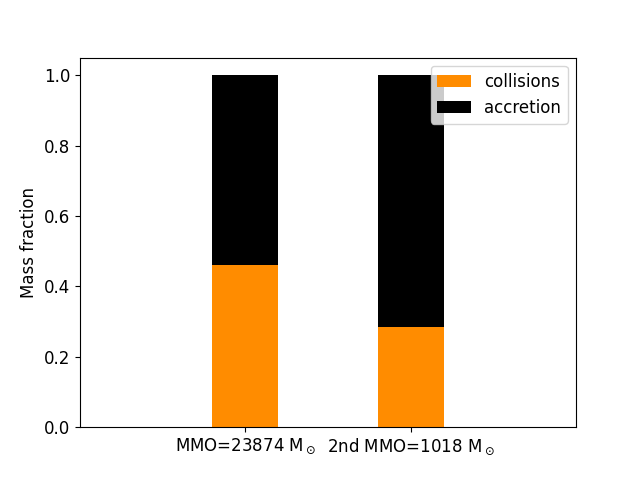}
    \caption{Average mass fractions gained through collisions and accretion, and average final masses for the MMO and the second MMO for simulations M3\_t100\_1--6.}
    \label{fig:fraction_collisions_m3e4_100}
\end{figure}

In all our simulations with $M_{\rm gas}=3\times10^4$~M$_\odot$ we see the formation of a single object that contains on average $80\pm7$ per cent of the initial cluster mass at 200~kyr.
This means that the average mass of the MMO is $23\,873\pm2\,001$~M$_\odot$.
We present in Fig.~\ref{fig:MMO_evol_m3e4} some of the properties of the MMO, like the mass, accretion rate, radius, and number of collisions it experiences during the evolution of the system.
This particle also evolves as a supermassive star due to the high accretion rates that it reaches, and grows both by accretion of gas and stellar collisions. The mass growth by mergers with other protostars contributes on average 46 per cent of its final mass as shown in Fig.~\ref{fig:fraction_collisions_m3e4_100}. Unlike in the less massive clusters, here higher accretion rates are reached, and they last for longer. We also see that stellar collisions contribute with a smaller mass fraction to the final mass of the MMO. This is simply due to the fact that in the simulations with $M_{\rm gas}=3\times10^4$~M$_\odot$, the MMO gains much more mass by gas accretion.

\begin{figure}
	\includegraphics[width=\columnwidth]{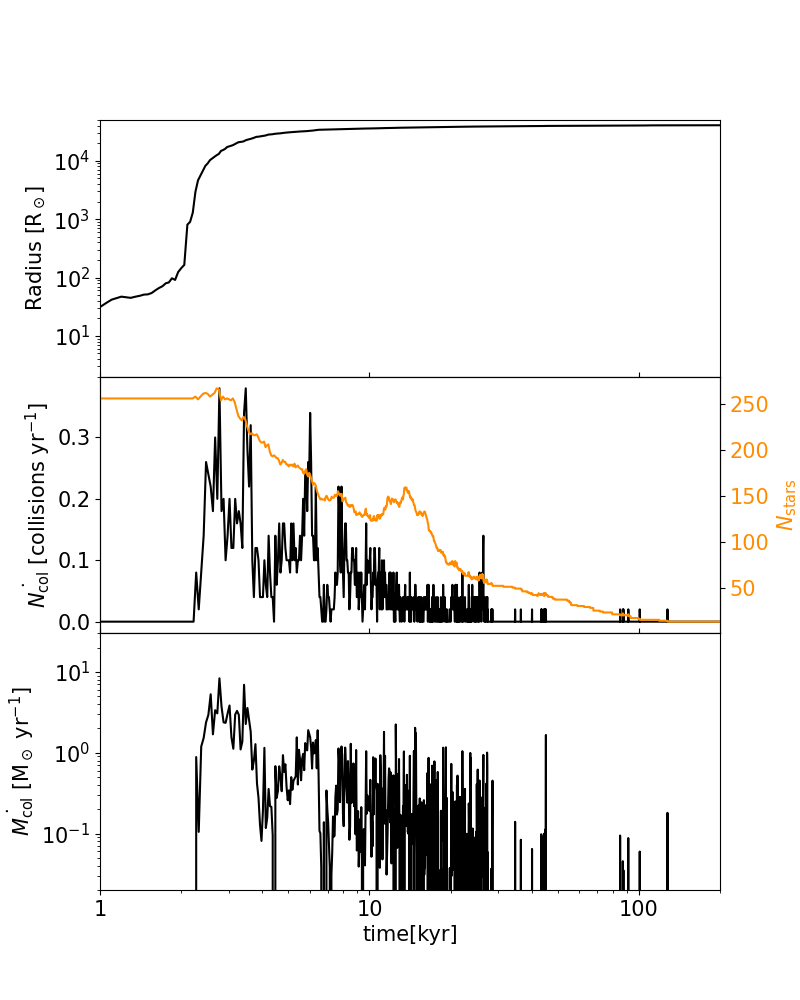}
    \caption{Same as Fig.~\ref{fig:col_rate_m1e4} but for a cluster with $M_{\rm gas}=3\times 10^4$~M$_\odot$.}
    \label{fig:col_rate_m3e4}
\end{figure}

Gas accretion peaks on average at 10~M$_\odot$~yr$^{-1}$ and remains above the critical accretion rate during the initial ~$8\, 000$~yr after the initial cloud contraction. This is sufficient to cause the protostar to evolve as an inflated object that quickly reaches a radius of more than 100~au, which in turn causes many stellar collisions to occur.
We see in Fig.~\ref{fig:col_rate_m3e4} that the collision rate peaks just after the MMO inflates in radius, reaching a peak of more than 0.3 collisions per year, a factor 3 higher than for the lower mass cloud simulations. The mass accretion due to collisions reaches peaks of $\sim10$~M$_\odot$~yr$^{-1}$, a factor 10 higher than for the less massive cloud simulations. The collision rate then decreases with the number of protostars.

\begin{figure}
	\includegraphics[width=\columnwidth]{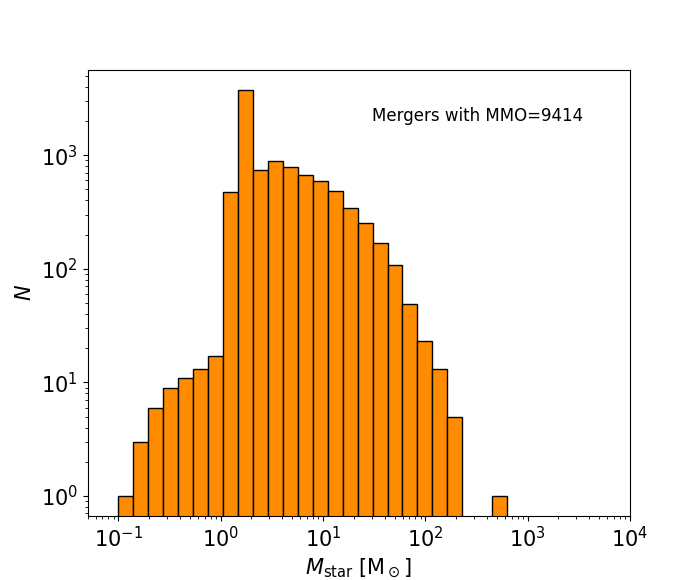}
    \caption{Mass distribution of the particles that merge with the MMO for simulation M3\_t100\_1.}
    \label{fig:masses_mergers_mmo_m3e4_100}
\end{figure}

Additionally, the huge gas densities ($\rho\sim10^{-8}$~g~cm$^{-3}$) found around the MMO trigger the formation of sink particles, $1\, 950$ new sink particles on average. Nearly all of these sink particles ($\sim 99\pm0.2$ per cent) merge with other objects, notably, $70\pm5$ per cent of the sinks merge with the MMO, and most of them do so shortly after they are created when they have accreted only 1--2~M$_\odot$. Sink particles in this mass range that merge with the MMO represent $52\pm9$ per cent of the total number of mergers, but they contribute on average only $11\pm4$ per cent of the total mass gained through mergers. We show the mass distribution of the sink particles that merge with the MMO in Fig.~\ref{fig:masses_mergers_mmo_m3e4_100}.

\subsubsection{Final cluster properties}
At the end of our simulations the remaining stellar systems consist of, on average, only $14\pm10$ particles with an average of $38\pm12$ ejected ones. Little to no gas is left and the final cluster is essentially comprised of an MMO with $\sim2\times10^4$~M$_\odot$ that is orbited by a few other stars, most of them with masses in the range 1--10~M$_\odot$. In 3 simulations, the second most massive object reaches more than $1\, 000$~M$_\odot$ and is orbiting the MMO in a close Keplerian orbit, but since the mass ratio $q=M_1/M_2$ is too high ($>20$), we do not mark them as binary systems.

\begin{figure}
	\includegraphics[width=\columnwidth]{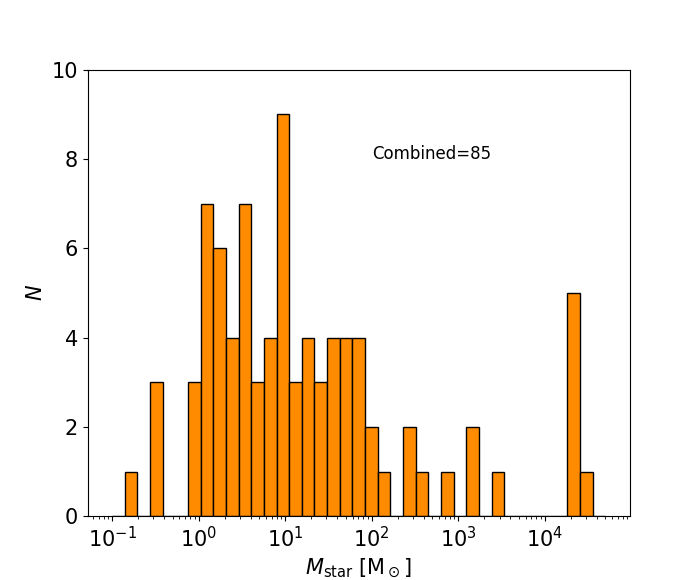}
    \caption{Combined mass distribution for stars bound to the cluster at the end of simulations M3\_t100\_1--6.}
    \label{fig:Combined_Mdist_m3e4_t100}
\end{figure}

We show the combined mass distribution of the particles that remain bound to the MMO for simulations M3\_t100\_1--6 in Fig.~\ref{fig:Combined_Mdist_m3e4_t100}. Comparing this mass distribution to the mass distribution of less massive clusters shown in Fig.~\ref{fig:mass_dist_m1e4}, the immediate difference that we note is that now we do not have a prominent peak. Instead  the mass distribution looks flat in the mass range 1--100~M$_\odot$.

\begin{figure}
	\includegraphics[width=\columnwidth]{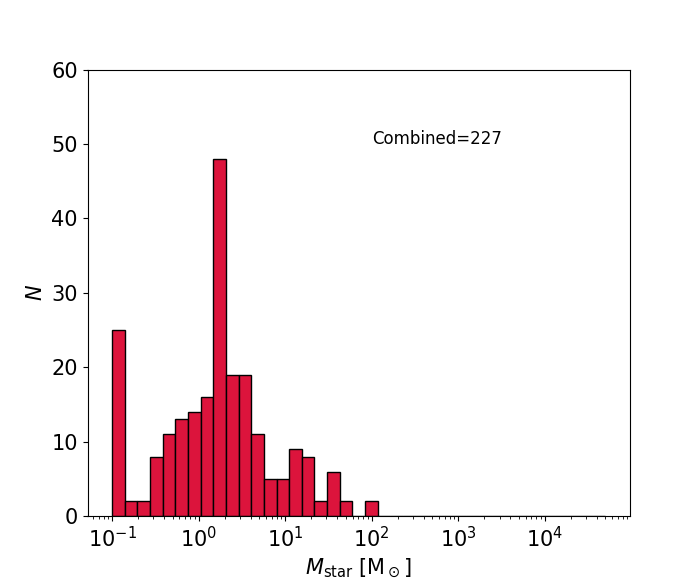}
    \caption{Combined mass distribution for stars ejected from the cluster at the end of simulations M3\_t100\_1--6.}
    \label{fig:Combined_Mdist_ejec_m3e4_t100}
\end{figure}

We also show the combined mass distribution of ejected particles for these simulations in Fig.~\ref{fig:Combined_Mdist_ejec_m3e4_t100}. This looks more similar to the one for less massive clusters, but with an additional peak at $\sim0.1$~M$_\odot$. While the shape of the mass function is maintained it now peaks in between 1--2~M$_\odot$ instead of 2--3~M$_\odot$ as observed for less massive clusters.\\

We present the individual mass distribution of bound and ejected particles for each of these simulations in Figs.~\ref{fig:M_dist_m3e4_individual}~and~\ref{fig:M_dist_m3e4_ejec_individual}, respectively.

\subsection{Impact of a reduced $t_{\rm KH, surf}$ }
\label{sec:rad_feed}
As explained at the end of Sec.\ref{sec:m-r_param} the time it takes for an inflated SMS to contract to the main sequence after its accretion rate falls below $\dot{M}_{\rm cirt}$ (i.e. $t_{\rm KH, surf}$) ranges in between 10--100~$t_{\rm KH}$ \citep{Sakurai2015}. So far we have assumed $t_{\rm KH, surf}=100~t_{\rm KH}$ and found that the MMOs in our simulations reach an average mass of $3\,729$~M$_\odot$ for clusters with $M_{\rm gas}=10^4$~M$_\odot$ and $23\,873$~M$_\odot$ for clusters with $M_{\rm gas}=3\times10^4$~M$_\odot$.


In principle a shorter $t_{\rm KH, surf}$ would cause an earlier contraction to the main sequence and this could impact the formation of an MMO via two effects. The first one is related to the growth via stellar collisions. An earlier contraction to the main sequence implies that the protostars that evolve as SMSs will not maintain a high cross-section for long enough times compared to simulations with $t_{\rm KH, surf}=100~t_{\rm KH}$. The second effect is a reduction of the mass of the stars that evolve as SMSs, given that gas accretion should terminate once the star contracts to the main sequence. Because of this, the simulations with $t_{\rm KH, surf}=10~t_{\rm KH}$ would have the highest impact on the final mass of the MMO.

In order to explore the effects of a reduced $t_{\rm KH, surf}$ on the final masses of the MMOs we ran simulations with $t_{\rm KH, surf}=10~t_{\rm KH}$ for our two different cluster models (see Table~\ref{tab:m1e4_outcomes}) and compare the results to the ones from our simulations with $t_{\rm KH, surf}=100~t_{\rm KH}$. This gives us an idea of how much a reduced $t_{\rm KH, surf}$ would impact the formation of an MMO through stellar collisions.
Then in order to get an idea of how our results would change when including radiation feedback we post process the snapshots of our simulations with $t_{\rm KH, surf}=10~t_{\rm KH}$. For this we stop the mass growth of all particles once they contract to the main sequence and use the existing merger history, but now with the modified masses to obtain a new estimate for the final mass of the MMO.


\subsubsection{Clusters with $M_{\rm gas}=10^4$~M$_\odot$}

For the set of simulations with $M_{\rm gas}=10^4$~M$_\odot$ (M1\_t100\_1--6 and M1\_t10\_1--6), by comparing the final masses of the MMOs (column 7) we find that the average values are consistent within one sigma errors. In fact, for simulations with $t_{\rm KH, surf}=100~t_{\rm KH}$, the average mass of the MMO is $\sim 3\,700\pm800$, whereas for simulations with $t_{\rm KH, surf}=10~t_{\rm KH}$, the average mass of the MMO is $\sim 4\,000\pm600$. 
Therefore we find that an earlier contraction to the main sequence has no impact on the growth of the MMO via stellar collisions and that the different values that we find here are the result of the intrinsic variability among different simulations.\\
The MMO does not contract to the main sequence but this is not due to very frequent stellar collisions, the mean time between collisions ($\sim2\,700$~yr) is actually slightly longer than 10~t$_{\rm KH}$ ($\sim1\,700$~yr). We attribute this behaviour to very short accretion bursts that surpass $\dot{M}_{\rm crit}$ during a brief period of time ($<50$~yr) not captured in Fig.~\ref{fig:MMO_evol_m1e4} since the cadence for data output is 50~yr.\\

Subsequently, after including an approximate effect of feedback (i.e., stopping the mass growth of a star once it enters the main sequence) we find very little reduction of the stellar masses. For simulations with $t_{
\rm KH, surf}=10~t_{\rm KH}$ the average final mass of the MMO goes down to $\sim3\,800\pm500$~M$_\odot$ after post-processing. This is still within the one sigma error of the value obtained without any type of feedback.
We show the comparison of the final masses of the MMOs in the case without feedback and in the case with approximate feedback in the left panel of Fig.~\ref{fig:app_feed}.\\

Finally, we also note that the same holds true for the binary systems formed in these simulations. We find a reduction of only 10 per cent for the masses of the primary and secondary stars, and the mass ratios remain the same.

\subsubsection{Clusters with $M_{\rm gas} = 3\times10^4$~M$_\odot$}
We first compare the final mass of the MMO for simulations with $t_{\rm KH, surf}=100~t_{\rm KH}$ and $t_{\rm KH, surf}=10 ~t_{\rm KH}$. For the first case the average mass of the MMO is $\sim23\,800\pm2\,000$~M$_\odot$, and for the second case the average mass of the MMO is $\sim22\,000\pm2\,200$~M$_\odot$. Thus we find again that an earlier contraction to the main sequence will not reduce the final mass of the MMO due to collisions appreciably. The MMO still remains inflated due to frequent mergers and this is the main driver of stellar collisions. The different values that we find here are a result of the intrinsic simulation to simulation variation.\\

\begin{figure}
	\includegraphics[width=\columnwidth]{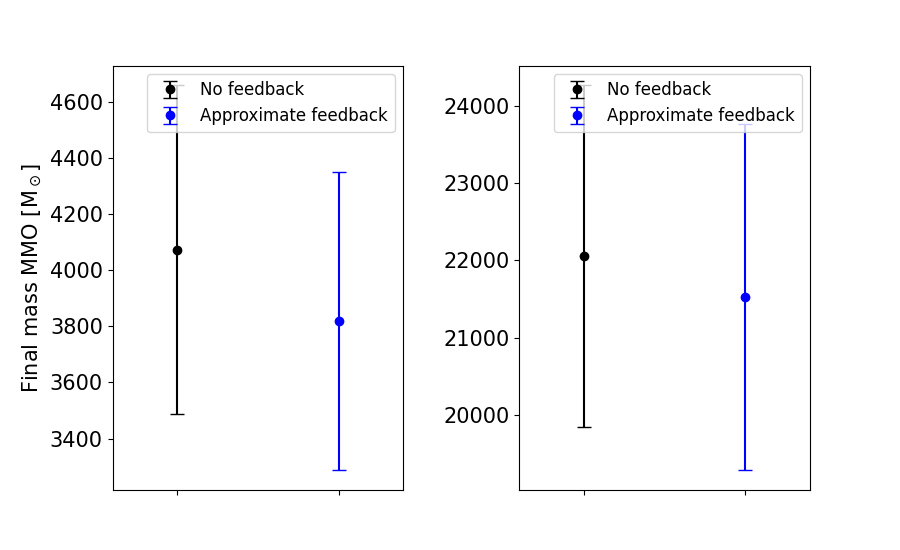}
    \caption{Estimate for the approximate effects of radiation feedback on the final mass of the MMO obtained from post-processing of our simulations. \textit{Left panel:} Average final mass of the MMO with one sigma errorbars for simulations with $M_{\rm gas}=10^4$~M$_\odot$ and $t_{\rm KH, surf}=10~t_{\rm KH}$. \textit{Right panel:} Same as left panel but for simulations with $M_{\rm gas}=3\times10^4$~M$_\odot$ and $t_{\rm KH, surf}=10~t_{\rm KH}$.}
    \label{fig:app_feed}
\end{figure}

When we post-process our simulations to account for the approximate effect of radiation feedback we find that the average mass of the MMO is reduced to $\sim21\,500\pm2\,200$~M$_\odot$ which again is within the one sigma error of the average mass of the MMO when no feedback is considered. We thus conclude that radiation feedback would not appreciably reduce the mass of the MMO. We show the final average masses of the MMOs when no feedback is included and with approximate feedback in the right panel of Fig.~\ref{fig:app_feed}.

We find however that the mass of the second most massive object is severely affected. On average the post-processing reduces the mass of the second MMO by a factor of four, and in one case even a factor of six. This occurs because in the simulations where a higher density is reached in the cloud centre, many protostars initially evolve as SMSs as they surpass the critical accretion rate. They however cannot maintain the high accretion rates for long times and eventually contract to the main sequence. In the simulations with high $t_{\rm KH, surf}$ this timespan of lower accretion rates is not long enough for the stars to contract to the main sequence , but in post-processing when $t_{\rm KH, surf}$ is lower they contract and their masses are fixed at that point.

\subsubsection{Possible impact of feedback from surrounding stars}
Regarding the formation and growth of the MMO we try to understand if once another protostar close to the MMO contracts to the main sequence, it would appreciably impact the final mass of the MMO. For this purpose we calculate the Str\"omgren radius of the first star that enters the main sequence in our simulations and that is close enough to influence the growth of the MMO.
We find that the first star enters the main sequence once the MMO has already gathered more than 80 per cent of its final mass, and the typical separation between these objects is around 500~au, but the Str\"omgren radius for the main sequence star (as calculated in Appendix~\ref{sec:Strom}) is usually around 6~au. Therefore we conclude that the surrounding stars that enter the main sequence will not appreciably impact the final mass of the MMO, which can also still grow through stellar mergers.

\section{Discussion and conclusions}
\label{sec:discussion}
In this work we address the problem of massive object formation, considering the evolution of a protostellar cluster embedded in a massive gas cloud at very low metallicity following fragmentation at sub-parsec scales, during the collapse of an atomic-cooling halo. We perform a suite of hydrodynamical plus $N$-body simulations including sink particle creation, gas accretion, pseudo (proto-)stellar evolution and stellar collisions. Our calculations start from a Plummer distribution of protostars embedded in a gas cloud that follows a Plummer density profile as well, and to which we impose a turbulent velocity field with Mach number equal to one, resembling the centre of a spherically collapsing atomic-cooling halo as found in cosmological hydrodynamical simulations \citep{Chon2018}. We note that spherical collapse is not always the case, and elongated clouds are also found in these simulations, so it will be desirable to begin with initial conditions taken directly from cosmological simulations in the future. Despite this, no fragmentation at parsec-scales is seen during these simulations, and we expect that fragmentation at smaller scales, as studied here, will not impede the formation of a supermassive star with 10$^4$~M$_\odot$. 

Our simulations include (proto)-stellar evolution in the sense that the stellar radius and luminosity change with the mass and accretion rate of the protostar. Therefore the protostars are allowed to contract to the main sequence and thus reduce the cross-section for collisions. This is essential because survival, ejection, and merger rates depend sensitively on the object size. One caveat in our simulations comes from the fact that we do not include stellar feedback. We estimate the possible impact of radiation feedback on the final mass of the MMO in Sec.~\ref{sec:rad_feed}. For this purpose we post-processed our simulations and stop mass accretion once the object enters the main sequence, at which point we assume that strong UV radiation may remove gas from the surrounding. In this extreme model, further mass growth is only possible via collisions. We find that this modification has negligible impact on the final masses of the MMO, and similar holds for the binary systems when present. We also calculate the Str\"omgren radius of surrounding stars that enter the main sequence but find that they do so once the MMO has already gathered more than 80 per cent of its final mass and these stars are not close enough to affect the growth of the MMO. We find however that the masses of the second most massive object could be affected in simulations with $M_{\rm gas}=3\times10^4$~M$_\odot$. This indicates that in order to fully characterize the final stellar masses in such systems, simulations that include radiation feedback are required.\\

The simulation results presented here agree with the study by \citet{Chon20} where they find what they termed a `super-competitive accretion' in which a single massive object dominates the growth by gas accretion. The conditions under which this scenario emerges were recently studied by means of analytical arguments by \citet{Schleicher2023}. They demonstrate that self-gravity induced accretion will initially dominate the mass growth of an object that will become the most massive object in the cluster and this does not depend on the number of protostars present. Only after the gas accretion mode shifts from self-gravity to Bondi-Hoyle, due to lower gas densities, could the fragments around the MMO interfere with gas accretion onto it, however, the moment at which this occurs depends only on the square root of the number of protostars. We conclude that in the context of atomic cooling halos as studied here, in an initially Jeans-unstable cloud, varying the initial number of protostars will have little to no impact on the mass growth through accretion.\\

We note that we model systems with a high initial number density of protostars (see Sec.~\ref{sec:init_cond}). We tried to estimate the effect that a lower initial number density of protostars would have on the final mass of the MMOs. For this purpose we consider the extreme scenario in which no initial protostars are present and post-process the collision histories to remove all the mass contributed by these protostars to the MMO. In this extreme scenario in which that mass would be lost from the system we find that for our simulations with $10^4$~M$_\odot$ in gas, the mass of the MMO decreases by 36 per cent; and in our simulations with $3\times10^4$~M$_\odot$ in gas, the mass is reduced by 11 per cent. This reduction in mass would still leave MMOs that can collapse to produce massive black holes seeds.\\

Since we consider protostars forming in a pristine gas cloud, the stars that are formed in our simulations resemble primordial stars, in particular the so-called Population III stars. These stars, once on the main sequence, do not lose significant mass due to stellar winds because of their low metallicity \citep{Krticka2006}, therefore including mass loss due to stellar winds will not change our results.

¿Moreover, the MMOs formed in our simulations evolve as supermassive stars because of the high accretion rate they reach \citep{Schleicher2013,Hosokawa13,Haemerle18b,Haemerle18a}. Again, due to their low metallicity, no mass loss is expected from stellar winds. Furthermore \cite{Hosokawa13} demonstrated that mass loss due to the pulsational instability reaches a maximum of $\sim5\times10^{-3}$~M$_\odot$~yr$^{-1}$, much lower than the accretion rates that these objects experience. It is thus safe to also ignore mass loss for our MMOs.
Note that mass loss due to stellar winds is relevant for higher metallicity stars when considering the formation of massive objects due to stellar collisions as explored in the context of nuclear star clusters by \citet{Das21}.

Finally, we note that the sink particles we consider resemble protostars but eventually some of them reach the zero age main sequence (ZAMS) and turn into stars. The point at which this typically occurs is around 30~kyr when protostars accreting at high rates reach the Kelvin–Helmholtz contraction phase \citep{Hosokawa09}. At this moment the MMO has gathered more than 80 per cent of its final mass. 
The typical mass for stars that reach the ZAMS is around 20~M$_\odot$. Stellar evolution models for these type of stars show main sequence lifetimes in the order of Myr even for very massive stars \citep{Tanikawa202,Murphy2021}, therefore no supernova explosion can occur during the time-span of our simulations.\\

We have not considered mass loss during stellar collisions. This effect has been studied in the context of blue straggler formation \citep{Sills1999,Sills2000}, and in the context of local star clusters. For this purpose, fitting functions depending on the mass ratio of the collision \citep{Lombardi2002} and the stellar structure \citep{Glebbeek2008,Glebbeek2008b} have been obtained for stars colliding at different stages during their evolution \citep{Glebbeek2013}. Armed with these functions \citet{Alister20} found that including mass loss could reduce the mass of the MMO by 20--40 per cent in a similar environment to the one studied here. Applied to our simulations, we conclude that the run-away formation of the MMO cannot be prevented. However, we note that it is not clear how well these analytical estimates can be applied to the collision between an SMS and its surrounding stars. More work is required to reduce the uncertainty in these estimates.\\

The formation of very massive objects that can collapse to produce massive black hole seeds has also been investigated in the context of star cluster formation in non-primordial clouds \citep{Sakurai2017,Sakurai2019,Das21}. In these models different mass-radius relations are used as the accretion rates experienced in these environments are much lower. In particular in these studies the stars never evolve as the inflated SMSs produced in our simulations.
This lead to important differences in the masses of the objects formed and the timescales involved. Even in presence of a much larger gas reservoir (10$^5$~M$_\odot$) the most massive objects reach typical masses of 10$^3$~M$_\odot$ \citep{Sakurai2019,Das21}.\\

According to previous simulations investigating the formation of SMSs in atomic cooling halos, mass inflows of 0.5~M$_\odot$~yr$^{-1}$ are reported at scales similar to the ones simulated here \citep{Wise2019}. At this constant rate the flow can be maintained for about 1~Myr which is comparable to the lifetime of the most massive stars formed in our simulations. Assuming that a total of 10$^6$~M$_\odot$ have concentrated in the inner 1~pc of the DM halo we estimate a binding energy of $\sim5\times10^{52}$~erg. On the other hand the binding energy of a 150~M$_\odot$ Population~III star (a typical massive star formed in our simulations) is around $1.7\times10^{52}$~erg. These estimates yield similar quantities, therefore it is very uncertain to say that a supernova explosion will or not be able to eject the remaining gas, it is important to know how much mass is concentrated inside which volume and the final masses of the stars.
If the gas is not ejected after the supernova explosion, another episode of star formation could occur in the halo but this time producing second generation stars due to the metal enrichment of the ejecta.\\

In our simulations we find that a massive central object is always formed and experiences run-away growth via collisions with other protostars in the cluster.The mass growth is typically dominated by one single object as found in previous studies \citep{Latif2013,Inayoshi2014,Sakurai2016,Matsukoba2019,Chon20} and explained by analytical arguments by \citet{Schleicher2023}.
Additionally the fragmentation process does not fully suppress the high mass flow towards the centre and so the MMO continues to grow via gas accretion as well. The MMO begins to grow once the cloud collapses on a free-fall timescale (around $3\,000$~yr) and by $10\,000$~yr it already contains 37 per cent of the initial gas mass for clusters with $M_{\rm gas}=10^4$~M$_\odot$, and 80 per cent of the initial gas mass for clusters with $M_{\rm gas}=3\times10^4$~M$_\odot$. 50 to 60 per cent of the mass of the MMO is gained through collisions.
In a third of the simulations with $M_{\rm gas}=10^4$~M$_\odot$ we find that the MMO is in a binary system with another massive object with mass ratios in between 1:2 and 1:7.
Radiation feedback is unable to reduce the mass of the MMO significantly. The final outcome is therefore a small group of tens of stars with typical masses in the range 1--100~M$_\odot$ orbiting a single object with $10^3$ or $10^4$~M$_\odot$. In one third of the cases the group of stars orbits a pair of massive objects ($\sim 10^3$~M$_\odot$) in a binary configuration.

\section*{Acknowledgements}
BR acknowledges support through ANID (CONICYT-PFCHA/Doctorado acuerdo bilateral DAAD/62180013) as well as support from DAAD (funding program number 57451854). 
%
PS acknowledges support through ANID/Doctorado en el Extranjero convocatoria 2022 (funding number 72220198).
The team acknowledges funding from the European Research Council via the ERC Synergy Grant ``ECOGAL'' (project ID 855130), from the Deutsche Forschungsgemeinschaft (DFG) via the Collaborative Research Center ``The Milky Way System''  (SFB 881 -- funding ID 138713538 -- subprojects A1, B1, B2 and B8) and from the Heidelberg Cluster of Excellence (EXC 2181 - 390900948) ``STRUCTURES'', funded by the German Excellence Strategy. We  also thank the German Ministry for Economic Affairs and Climate Action for funding in the project ``MAINN'' (funding ID 50OO2206). DRGS gratefully acknowledges support by the ANID BASAL projects ACE210002 and FB210003, via the Millenium Nucleus NCN19-058 (TITANs) and via Fondecyt Regular (project code 1201280).
Part of the simulations presented in this work  were performed with resources provided by the Kultrun Astronomy Hybrid Cluster via the projects Conicyt Programa de Astronomia Fondo Quimal 2017 QUIMAL170001, Conicyt PIA ACT172033, and Fondecyt Iniciacion 11170268. Most of the simulations were performed on the Helix cluster and BwUniCluster2.0 supported by the state of Baden-W\"{u}rttemberg through bwHPC and the German Research Foundation (DFG) through grant INST 35/1597-1 FUGG, data are stored at SDS@hd funded through grant INST 35/1314-1 FUGG. 

\section*{Data Availability}
The data underlying this article will be shared on reasonable request to the corresponding author.
 



\bibliographystyle{mnras}
\bibliography{example} 




\appendix

\section{Mass radius parametrization}
\label{sec:app_mr}
We use different mass-radius (M-R) relations depending on the accretion rate of the protostar and on its evolutionary stage. We also calculate associated quantities such as the luminosity and the Kelvin Helmholtz (KH) timescale. All these properties are calculated after every accretion step. We define three evolutionary stages, namely \textit{protostar}, \textit{star}, and \textit{supermassive star}. The M-R relations for each stage are described below.

\subsection{\textit{Protostar}}
Every particle in our simulations begins in the \textit{protostar} stage. The M-R parametrizations that we use for them are based on the works of \cite{Hosokawa09} and \citet{Hosokawa12,Hosokawa13}. We calculate the properties of each \textit{protostar} by classifying them into three different tracks. The classification depends on the accretion rate $\dot{M}$. We therefore have the `SMS' track, `VMS' track and `NORMAL' track. Each track is described in the next subsections.

\subsubsection{`SMS' track}
There is a critical accretion rate above which the accreting protostars remain inflated and their radii always increase with the mass. A \textit{protostar} whose accretion rate is higher than this critical accretion rate is in the `SMS' track. The critical accretion rate in our simulations is set to $\dot{M}_{\rm crit} = 0.04$~M$_\odot$~$yr^{-1}$ taken from \cite{Hosokawa13}. 
For every \textit{protostar} in the `SMS' track, the radius is computed as:
\begin{equation}
    \label{eq:M_R_SMS_app}
    R_* = 2\,600 \left( \frac{M_*}{100~{\rm M_\odot}}\right)^{1/2} {\rm R_\odot}.    
\end{equation}

The \textit{protostar} will follow this relation unless the accretion rate $\dot{M}$ remains below $\dot{M}_{\rm crit}$ for more than 10--100~$t_{\rm KH}$ \citep{Sakurai2015,Schleicher2013}, where the KH timescale $t_{\rm KH}$ is calculated as:
\begin{equation}
    t_{\rm KH} = \frac{G M^2}{RL},
\end{equation}
with $M$, $R$, and $L$, being the mass, radius and luminosity of the protostar.  
For the calculation of the KH timescale we need the luminosity of the \textit{protostar}. As long as the mass is $\leq$~10~M$_\odot$ the luminosity is calculated as \citep{Hosokawa09}:
\begin{equation}
   \label{eq:L_ad}
    L_* = 0.6\left( \frac{M_*}{{\rm M_\odot}} \right)^{11/2} \left( \frac{R_*}{{\rm R_\odot}} \right)^{-1/2}~{\rm L_\odot} ,
\end{equation}
whereas for $M_*>10$~M$_\odot$, the luminosity is given by:
\begin{equation}
   \label{eq:L_kh}
    L_* = 10 \left( \frac{M_*}{M_\odot} \right)^{3}~{\rm L_\odot} ,
\end{equation}
and for $M_*>70$~M$_\odot$, the luminosity approaches the Eddington limit and is calculated as:
\begin{equation}
   \label{eq:L_edd}
    L_* = 3.8\times 10^6 \left( \frac{M_*}{100~{\rm M_\odot}} \right)~{\rm L_\odot}.
\end{equation}

Finally, once the \textit{protostar} reaches a mass of 600~M$_{\odot}$, it enters the \textit{supermassive star} stage. In case the accretion rate $\dot{M}$ of a \textit{protostar} in the `SMS' track remains below $\dot{M}_{\rm crit}$ for more than a 10--100 KH timescales, the \textit{protostar} will enter a new evolutionary track according to its last value for $\dot{M}$. We note that the time during which a protostar in the `SMS' track remains inflated after its accretion rate falls below $\dot{M}_{\rm crit}$ can vary between 10--100 KH timescales \citep{Sakurai2015}. We consider both extreme values for the KH timescales in this work.

\subsubsection{`VMS' track}
Every \textit{protostar} whose accretion rate is in the range $[10^{-6},0.04[$~M$_\odot$~yr$^{-1}$ is in the `VMS' track. In this track we distinguish three phases, the \textit{adiabatic accretion phase}, the \textit{swelling}, and the \textit{Kelvin Helmholtz contraction} as described in \cite{Hosokawa09}. The \textit{adiabatic accretion phase} holds as long as the mass of the \textit{protostar} is $\leq$~$M_{\rm ad}$ which is given by:
\begin{equation}
    \label{eq:m_ad}
    M_{\rm ad} = 0.9  \left[ \left(\frac{\dot{M}}{4.2\times 10^{-8} {\rm M_\odot yr^{-1}}}\right) \left(\frac{\dot{M}}{10^{-3} {\rm M_\odot yr^{-1}}}\right)^{(-0.41/2)} \right]^{(2/9.27)} \rm M_\odot.
\end{equation}
In this phase the luminosity is calculated from Eq.(\ref{eq:L_ad}) and the radius as:
\begin{equation}
    R_*=26 \left(\frac{M_*}{\rm M_\odot}\right)^{0.27} \left(\frac{\dot{M}}{10^{-3} {\rm M_\odot ~yr^{-1}}}\right)^{0.41} \rm R_\odot.
\end{equation}
Now it is useful to define two parameters $\alpha$ and $\beta$:
\begin{equation}
    \alpha=26 \left(\frac{M_{\rm ad}}{\rm M_\odot}\right)^{-4.73} \left(\frac{\dot{M}}{10^{-3} {\rm M_\odot ~yr^{-1}}}\right)^{0.41}, 
\end{equation}
\begin{equation}
    \beta=\alpha \left(1.2 \frac{M_{\rm ad}}{\rm M_\odot}\right)^{7.5}.
\end{equation}

The \textit{swelling} phase holds for \textit{protostars} whose mass is in the range $[M_{\rm ad},1.2M_{\rm ad}[$. In this phase the luminosity is given by Eq.(\ref{eq:L_ad}) and the radius as:
\begin{equation}
    R_*=\alpha \left(\frac{M_*}{\rm M_\odot}\right)^5 \rm R_\odot.
\end{equation}

The \textit{Kelvin Helmholtz contraction} phase holds for \textit{protostars} whose mass is in the range $[1.2M_{\rm ad},M_{\rm ms}[$, with $M_{\rm ms}$ given by:
\begin{equation}
    M_{\rm ms} = \left( \frac{\beta}{0.97}\right)^{1/3.07}.
\end{equation}
The luminosity in this phase is given by Eq.(\ref{eq:L_kh}) and the radius by:
\begin{equation}
    R_*=\beta \left(\frac{M_*}{\rm M_\odot}\right)^{-2.5} \rm R_\odot.    
\end{equation}

If the mass of a \textit{protostar} in the `VMS' track is larger than $M_{\rm ms}$, it enters the \textit{star} stage, i.e., it contracts to the main sequence. In case the accretion rate $\dot{M}$ of a \textit{protostar} in the `VMS' track remains below $10^{-6}$~M$_{\odot}$~yr$^{-1}$ for more than a KH timescale, the \textit{protostar} will enter the `NORMAL' track.

\subsubsection{`NORMAL' track}
The \textit{protostars} whose accretion rate is $\dot{M}<10^{-6}$~M$_{\odot}$~yr$^{-1}$ are in the `NORMAL' evolutionary track. As long as their mass is $<0.8$~M$_\odot$, the luminosity is given by Eq.(\ref{eq:L_ad}) and their radius is given by:
\begin{equation}
    R_*=0.86 \left( \frac{M_*}{\rm M_\odot} \right)^{0.27} \rm R_\odot. 
\end{equation}

If the mass is $\geq$~0.8~M$_\odot$ the \textit{protostar} enters the \textit{star} stage.

As long as a particle is in the \textit{protostar} stage, it can move between the three tracks described above. There is a hierarchy for the tracks, with the `SMS' track having the highest hierarchy and the `NORMAL' track having the lowest hierarchy. A \textit{protostar} can move to a track with higher hierarchy if the accretion rate $\dot{M}$ is high enough, but it will only move to a lower hierarchy track if the accretion rate $\dot{M}$ remains lower than the critical accretion rate for the current track during a time longer than a KH timescale.

\subsection{\textit{Star}}
When a particle is in the \textit{star} stage, the radius and luminosity are those for a star in the main sequence.
Therefore the luminosity is given by Eq.(\ref{eq:L_edd}) and the radius is calculated as:
\begin{equation}
\label{eq:ms_mr}    
    R_*=0.97 \left(\frac{M_*}{\rm M_\odot}\right)^{0.57} \rm R_\odot.
\end{equation}
We assume that a particle in the \textit{star} stage will not inflate in radius even if the accretion rate is $\dot{M}\geq0.04$~M$_\odot$~yr$^{-1}$, therefore once a star particle has entered the \textit{star} stage, it will always follow the same M-R relation.

\subsection{\textit{Supermassive star}}
When a particle is in the \textit{supermassive star} stage, the luminosity is given by Eq.(\ref{eq:L_edd}) and the radius is given by Eq.(\ref{eq:M_R_SMS_app}). It is worth noting that
even after a star particle has entered the \textit{supermassive star} stage, it can still contract if the accretion rate falls below the critical accretion rate for more than 10--100~KH~timescales. If that occurs, then the star particle will enter the \textit{star} stage and will follow the mass radius relation given by Eq.~\ref{eq:ms_mr}.

\section{Properties of the merger product}
\label{sec:properties_merger}
In our simulations the star particles have not only mass and radius, but also
an interaction zone radius, angular momentum, an evolutionary stage, and track (see Appendix~\ref{sec:app_mr}). We describe here the method that we follow to determine the
new properties for the merger product.

We make the assumption that the mass is conserved and the mass of the merger product is the sum of the masses of the progenitors. We do the same for the angular momentum. For the radius of the interaction zone we select the largest value among the two progenitors.

\subsection{Stage}
Determining the stage of the merger product is important to decide the new radius.
Given that particles can be in three different stages, we have six combinations
for the merging particles as presented in Table~\ref{tab:new_stage}.

\subsection{Track}
When the stage of the merger product is decided to be a \textit{protostar}, there are six possibilities for the \textit{track} in which
it can be, as shown in Table~\ref{tab:new_track}.
\begin{table}
	\centering
	\caption{Determination of the evolutionary stage for a merger product}
	\label{tab:new_stage}
	\begin{tabular}{l l l l} 
		\hline
		progenitor's stage & \textit{protostar} & \textit{star} & \textit{supermassive star} \\
		\hline
\textit{protostar} & \textit{protostar} & \textit{protostar} & \textit{supermassive star} \\

\textit{star} &\textit{-} & \textit{star} & \textit{supermassive star} \\

\textit{supermassive star} & \textit{-} & \textit{-} & \textit{supermassive star} \\

		\hline
	\end{tabular}
\end{table}

\begin{table}
	\centering
	\caption{Determination of the evolution track for a merger product in the \textit{protostar} stage}
	\label{tab:new_track}
	\begin{tabular}{l l l l} 
		\hline
		progenitor's track & NORMAL & VMS & SMS \\
		\hline
NORMAL & NORMAL & VMS & SMS \\

VMS & - & VMS & SMS \\

SMS & - & - & SMS \\

		\hline
	\end{tabular}
\end{table}

If the merging particles happen to be in the \textit{protostar} and \textit{star} stage, then the track of the merger product is simply the track of the \textit{protostar}.

For deciding the radius of the merger product we need to determine the accretion rate. We do this by assigning the highest accretion rate among the progenitors to the merger product.
Once the stage, the track, and the accretion rate of the merger product have been decided, the radius is calculated as described in Sec.~\ref{sec:app_mr}, and a new luminosity and KH timescale are also computed.

\section{Str\"omgren radius calculation}
\label{sec:Strom}
In order to provide an estimate for the impact of the stars that contract to the main sequence during the cluster evolution, we provide our calculations of the Str\"omgren radius at the moment when a star enters the main sequence for the first time in one of our simulations, specifically simulation M1\_t100\_1.

The Str\"omgren radius is given by
\begin{equation}
    R_s = \left( \frac{3}{4\pi} \frac{Q_\star}{n_H^2 \beta_2(T)}\right)^{1/3},
\end{equation}
where $Q_\star$ is the number of hydrogen ionizing photons ($h\nu \geq 13.6$~eV) per unit time, $n_H$ is the hydrogen nuclei number density and $\beta_2(T)$ is the case B volume recombination rate for hydrogen, which is $\sim2\times10^{-13}$~cm$^{-3}$~s$^{-1}$ at 10$^4$~K. We adopt this value here.

The number of ionizing photons can be estimated by using Planck's function as
\begin{equation}
 \label{eq:photon_rate}
    Q_\star = \int_{\nu_1}^{\infty} \frac{L_\nu}{h\nu} d\nu= \frac{8\pi^2 R^2}{c^2} \int_{\nu_1}^{\infty} \frac{\nu^2}{{\rm e}^{\frac{h\nu}{kT}} -1 }, 
\end{equation}
where $h\nu_1=13.6$~eV and $R$ is the radius of the star. 
Now, in this specific simulation the masses of the first star that enters the MS is around 7.5~M$_\odot$, this implies a radius of $R\sim3$~R$_\odot$, luminosity of $\sim 285\,363$~L$_\odot$, and $T_{\rm eff}\sim 76\,316$~K. With these values we obtain $\frac{h\nu_1}{kT}\sim 2.1$, so we can approximate the integral in eq.(\ref{eq:photon_rate}) as
\begin{equation}
    \int_{\nu_1}^{\infty} \frac{\nu^2}{{\rm e}^{\frac{h\nu}{kT}} -1 } d\nu \sim \int_{\nu_1}^{\infty} \frac{\nu^2}{ {\rm e}^{\frac{h\nu}{kT} } } d\nu = \left( \frac{kT}{h} \right)^3 \int_{x_1}^{\infty} \frac{x^2}{ {\rm e}^x } dx 
\end{equation}

\begin{equation*}
   \ \ \ \ = \left( \frac{kT}{h} \right)^3 (x_1^2 + 2x_1 + 2){\rm e}^{-x},
\end{equation*}
where $x=\frac{h\nu}{kT}$.

By doing so, we obtain
\begin{equation}
   Q_{\star}=\frac{8\pi^2 R^2}{c^2} \left( \frac{kT}{h} \right)^3 \left[ \left(\frac{h\nu_1}{kT}\right)^2 + 2\frac{h\nu_1}{kT} +2  \right] {\rm e}^{-\left(\frac{h\nu_1}{kT}\right)}.
\end{equation}
Inserting the numerical values we obtain
\begin{equation}
   Q_{\star}\sim2\times10^{49}~{\rm s^{-1}}.
\end{equation}

For estimating the number density of hydrogen atoms, we take the mean density inside a sphere with radius 200~au, centered on the radiation source, and obtain $n_H=5.6\times10^9$~cm$^{-3}$, so the Str\"omgren radius is 
\begin{equation}
    R_s=6~\rm au.
\end{equation}

We note that the separation between this star and the accreting MMO is 734~au at this moment, thus this star would be unable to stop further accretion onto the MMO.

\section{Mass distributions}
In this appendix we present the mass distribution of bound and ejected stars, at the end of, and for each of our simulations with $t_{\rm KH, surf}=100~t_{\rm KH}$.
\subsection{Clusters with $M_{\rm gas}=10^4$~M$_\odot$}
\begin{figure*}
	\includegraphics[width=\textwidth]{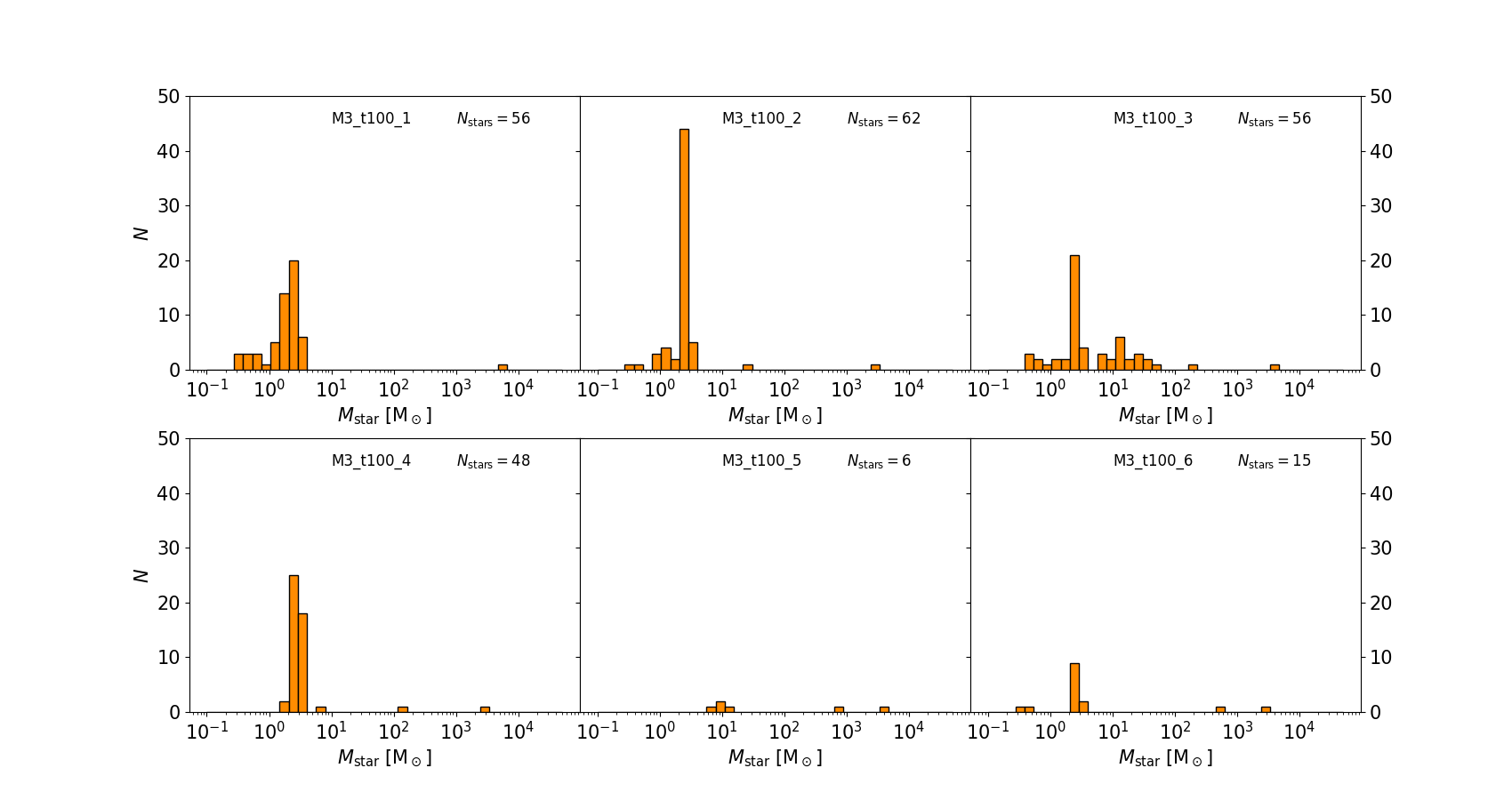}
    \caption{Mass distribution of bound stars at the end of our simulations M1\_t100\_1--4.}
    \label{fig:M_dist_m1e4_individual}
\end{figure*}

\begin{figure*}
	\includegraphics[width=\textwidth]{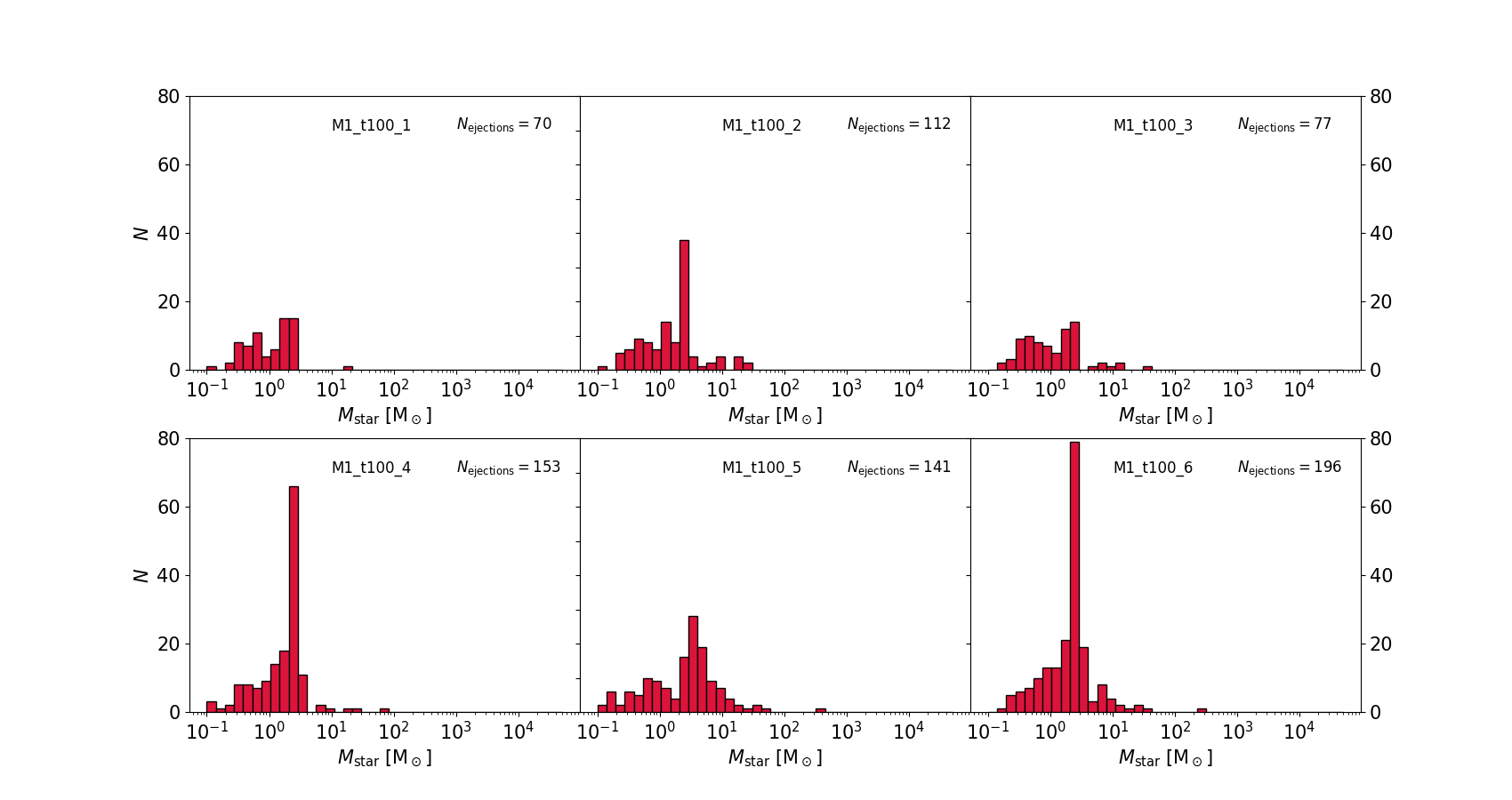}
    \caption{Mass distribution of ejected stars at the end of our simulations M1\_t100\_1--4.}
    \label{fig:M_dist_esc_m1e4_individual}
\end{figure*}

\subsection{Clusters with $M_{\rm gas}=3\times10^4$~M$_\odot$}
\begin{figure*}
	\includegraphics[width=\textwidth]{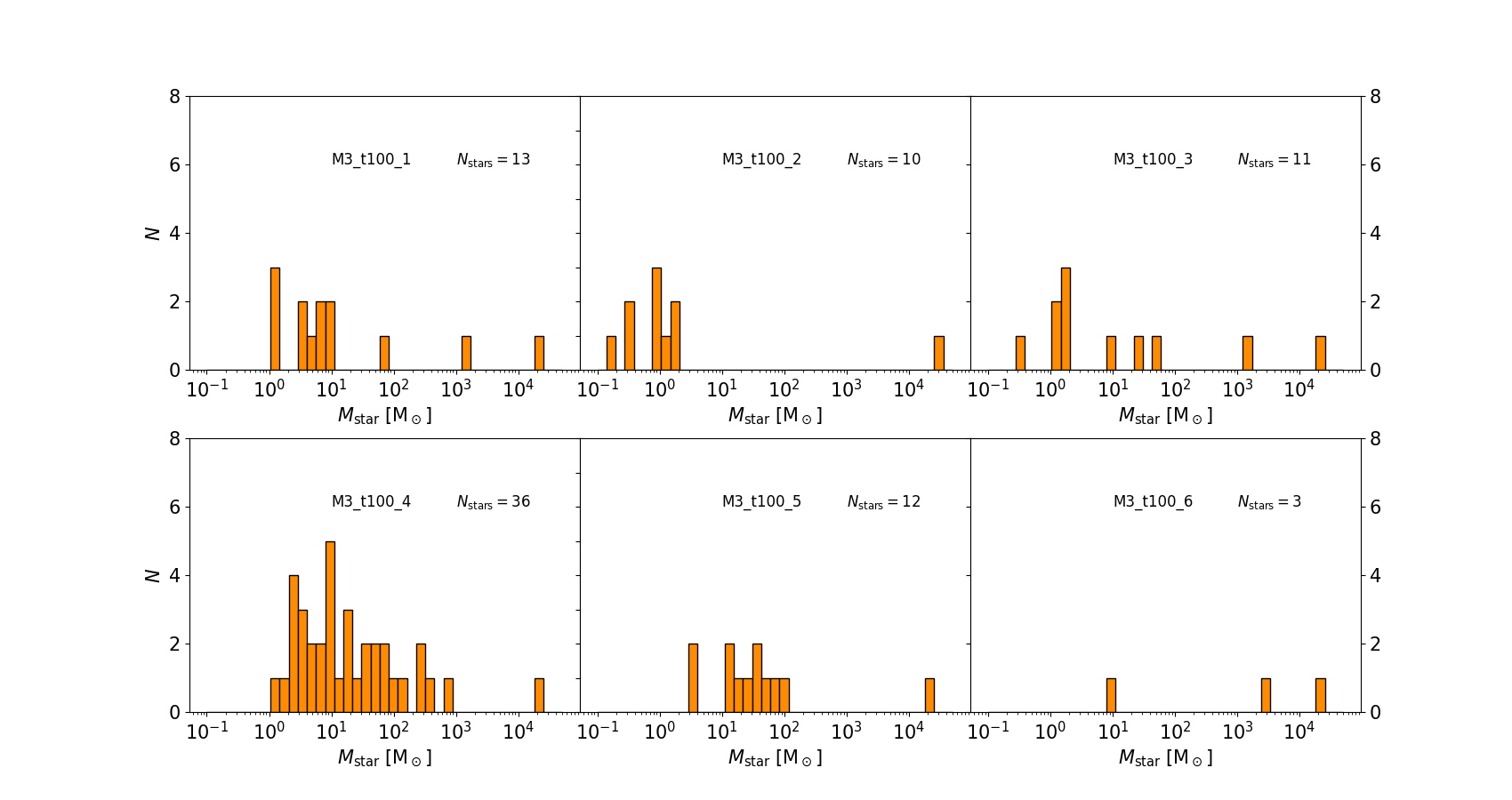}
    \caption{Mass distribution of bound stars at the end of our simulations M3\_t100\_1--6.}
    \label{fig:M_dist_m3e4_individual}
\end{figure*}

\begin{figure*}
	\includegraphics[width=\textwidth]{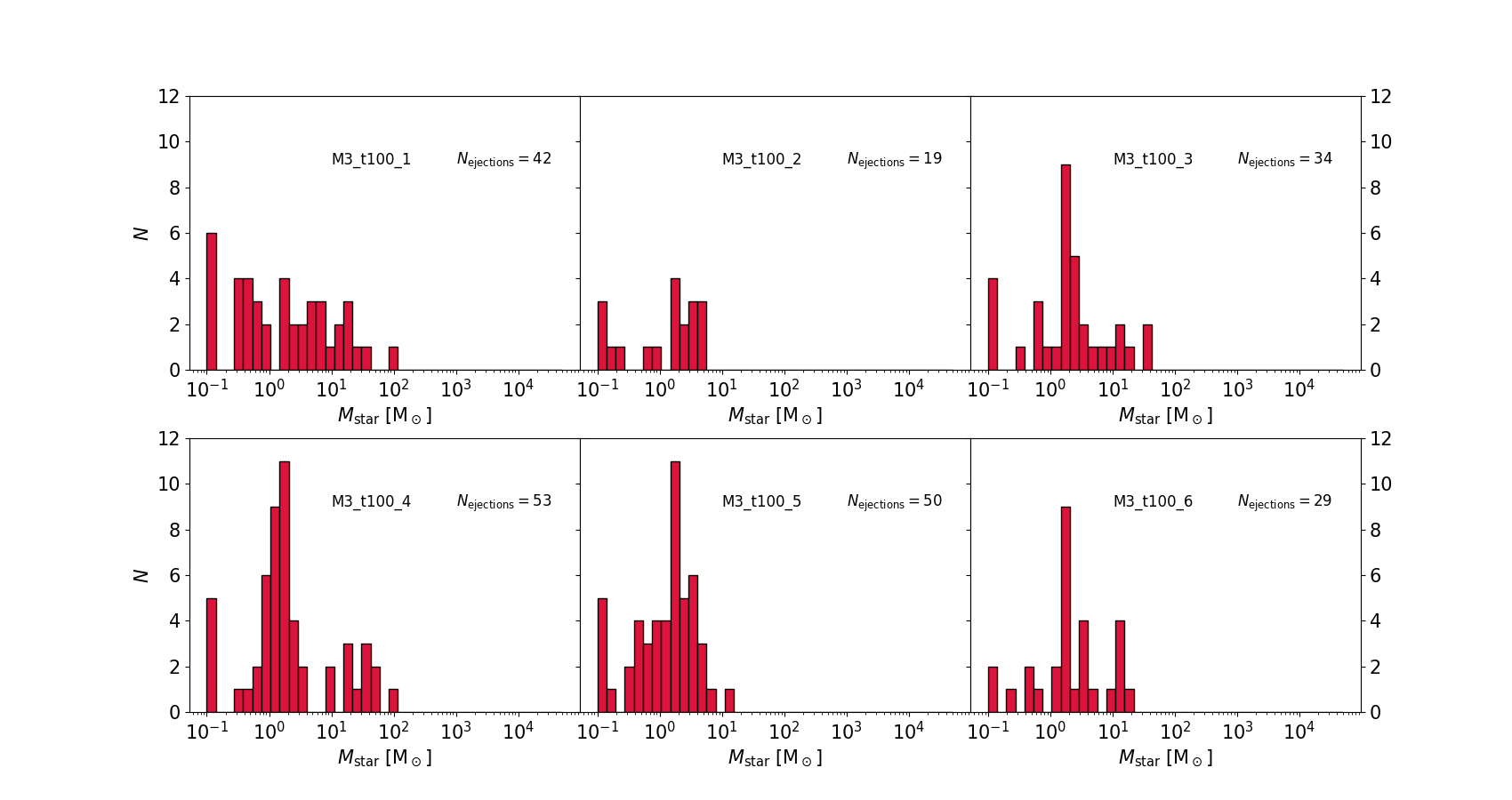}
    \caption{Mass distribution of ejected stars at the end of our simulations M3\_t100\_1--6.}
    \label{fig:M_dist_m3e4_ejec_individual}
\end{figure*}


\bsp	
\label{lastpage}
\end{document}